\documentclass[useAMS,usenatbib]{mn2e}
\usepackage{graphicx}
\usepackage{amssymb}
\usepackage{amsmath}
\usepackage{subfigure}
\usepackage{ifpdf}

\def\aap{A\&A}
\def\apjl{ApJL}
\def\apjs{ApJS}

\title[ISM Turbulence and Dynamical Heating]
  {A unified model for age-velocity dispersion relations in Local Group galaxies: Disentangling ISM turbulence and latent dynamical heating}
\author[Leaman et al.]
{Ryan Leaman$^{1}$\thanks{Email: leaman@mpia.de}, J. Trevor Mendel$^{2}$, Emily Wisnioski$^{2}$, Alyson M. Brooks$^{3}$,  
\newauthor Michael A. Beasley$^{4,5}$, Else Starkenburg$^{6}$, Marie Martig$^{1}$, Giuseppina Battaglia$^{4,5}$, 
\newauthor Charlotte Christensen$^{7}$, Andrew A. Cole$^{8}$, T. J. L. de Boer$^{9}$, and Drew Wills$^{7}$\\
$^{1}$Max-Planck Institut f\"ur Astronomie, K\"onigstuhl 17, D-69117 Heidelberg, Germany\\
$^{2}$Max-Planck Institut f\"ur Extraterrestrische Physik, Giessenbachstr. 1, D-85741 Garching, Germany\\
$^{3}$Department of Physics \& Astronomy, Rutgers, The State University of New Jersey, 136 Frelinghuysen Rd, Piscataway, NJ 08854, USA\\
$^{4}$Instituto de Astrof\'isica de Canarias, V\'ia L\'actea s/n 38205 La Laguna, Spain\\
$^{5}$Departamento de Astrof\'isica, Universidad de La Laguna, 38206 La Laguna, Spain\\
$^{6}$Leibniz-Institut für Astrophysik Potsdam, An der Sternwarte 16, D-14482 Potsdam, Germany\\
$^{7}$Grinnell College Physics Department, Noyce Science Center, Grinnell, IA 50112, USA\\
$^{8}$School of Physical Sciences, University of Tasmania, Hobart, TAS 7001, Australia\\
$^{9}$Institute of Astronomy, University of Cambridge, Madingley Road, Cambridge CB3 0HA, UK\\
}
\date{Accepted 2047 March 23. Received 2047 February 9; in original form 1993 December 9}

\pagerange{\pageref{firstpage}--\pageref{lastpage}} \pubyear{2014}

\begin{document}

\label{firstpage}

\maketitle

\begin{abstract}
We analyze age-velocity dispersion relations (AVRs) from kinematics of individual stars in eight Local Group galaxies ranging in mass from Carina ($M_{*} \sim 10^{6} M_{\odot}$) to M31 ($M_{*} \sim 10^{11} M_{\odot}$).  Observationally the $\sigma$ vs. stellar age trends can be interpreted as dynamical heating of the stars by GMCs, bars/spiral arms, or merging subhalos; alternatively the stars could have simply been born out of a more turbulent ISM at high redshift and retain that larger velocity dispersion till present day - consistent with recent IFU kinematic studies. To ascertain the dominant mechanism and better understand the impact of instabilities and feedback, we develop models based on observed SFHs of these Local Group galaxies in order to create an evolutionary formalism which describes the ISM velocity dispersion due to a galaxy's evolving gas fraction.  These empirical models relax the common assumption that the stars are born from gas which has constant velocity dispersion at all redshifts. Using only the observed SFHs as input, the ISM velocity dispersion and a mid-plane scattering model fits the observed AVRs of low mass galaxies without fine tuning.  Higher mass galaxies above $M_{vir} \gtrsim 10^{11} M_{\odot}$ need a larger contribution from latent dynamical heating processes (for example minor mergers), in excess of the ISM model. Using the SFHs we also find that supernovae feedback does not appear to be a dominant driver of the gas velocity dispersion compared to gravitational instabilities - at least for dispersions $\sigma \gtrsim 25$ km s$^{-1}$.  Together our results point to stars being born with a velocity dispersion close to that of the gas at the time of their formation, with latent dynamical heating operating with a galaxy mass-dependent efficiency.  These semi-empirical relations may help constrain the efficiency of feedback and its impact on the physics of disk settling in galaxy formation simulations.
\end{abstract}


\section{Introduction}
The dynamical evolution of galactic stellar disks is encoded in present day observations of the kinematics of long lived stars.  In these systems the rotation velocity and velocity dispersion ($\sigma$) of the stars provide a means to infer the mass distribution of the galaxy \citep{Cappellari09}, and quantify the angular momentum and orbital distribution of the stellar component \citep{Remco08}.  Understanding how the angular momentum of the stellar disk evolves over time offers a window to look at processes which may be responsible for redistributing stars within the galaxy, shaping its present day dynamical structure. Such dynamical diagnostics offer complementary insight to structural studies, for example the correlations between disk thickness and galaxy mass seen in studies of edge-on galaxies \citep{Kregel02}, or locally the anti-correlation between scale height and length seen amoungst the (thick) disk(s) of different ages in the MW \citep{Bovy12,Minchev15}.

The primary observational metric for studying the dynamical evolution of resolved systems is to compare the velocity dispersion as function of stellar age - the age velocity dispersion relation (AVR).  Measurements of the AVR from local solar neighbourhood stars in the MW have long provided a detailed chance to study local dynamical heating processes \citep{SS51,SS53,Wielen77}, the effect of mergers on the disk structure \citep{TothOst92,Hopkins08,Helmi12} and radial migration (e.g., \citealt{SellwoodBinney02,Roskar08,Loebman11,Grand15}).  The increase in $\sigma$ for the oldest stars in the solar neighbourhood AVR \citep{Holmberg09,Casagrande11} suggests that over time a scattering mechanism has dynamically heated the stellar populations - with the oldest stars showing higher velocity dispersion due to their longer exposure to this scattering from giant molecular clouds (GMCs) or spiral arms. 

These processes may work proportionally more in the vertical or in-plane directions, which for the Solar neighbourhood where the velocity dispersion in three components ($\sigma_{z},\sigma_{\phi},\sigma_{R}$) is measurable, provides an exquisite opportunity to understand the efficiency of these mechanisms through simulations (e.g., \citealt{House11,Grand16,Gustafsson16}; though this can be complicated due to volume selection effects, \citealt{Aumer16}) and their dependencies with other galactic properties such as the shape of the rotation curve (e.g., \citealt{Ida93,Ida99}).  
 
With the advent of high throughput near-infrared integral field units (IFUs) on 8m class telescopes, complementary observations of the ionized (and molecular) gas velocity dispersion in high redshift galaxies have become possible.  A high fraction of these galaxies have been shown to be rotation dominated disk galaxies in their kinematics and from deep NIR imaging. Recent studies have found that galactic disks exhibit much more turbulent interstellar mediums (ISMs) at higher redshifts \citep{Forster09,Genzel11,Saintonge11,Kassin12,Stott16}.  If the stars were initially born with the kinematic signatures ($\sigma_{ISM}$) of their gaseous birthplaces, we would then also expect that the oldest surviving stars in a galaxy have higher velocity dispersions than the younger stars. 

The exact nature of the mechanism(s) that power the large gas velocity dispersions in high redshift galaxies is still actively debated.  Momentum injection to the ISM from SNe explosions is theoretically able to provide turbulent support to the ISM, which in turn regulates the star formation rate within gaseous disks (c.f., \citealt{Ostriker11}).  Observationally, \cite{Lehnert13} suggested that the dispersion fields in high redshift galaxies were consistent with being generated by this SNe driven turbulence - although with a SNe energy injection efficiency which is much lower than what is commonly needed in simulations to reproduce dynamical, chemical and structural scaling relations \citep{Agertz13,Shen14,dicintio14}.  An alternative driver of velocity dispersion in the ISM is simply that large gas fractions in high redshift gas disks \citep{Daddi10} lead to instabilities which subsequently fragment the disk into massive clumps and gravitationally provide a source of random motion in the ISM \citep{Bournaud10}.

Understanding the relative contribution of these two internal processes is crucial given that high SNe feedback efficiencies appear necessary to resolve many issues in galaxy formation: from the apparent deviations from NFW halo profiles in low mass galaxies \citep{Governato10,Brooks14}, to the galactic mass-metallicity relation \citep{Dave12}, and abundances of gas in the IGM and CGM (e.g., \citealt{Shen13}).  Disk instabilities are equally important as an evolutionary mechanism, as the high mass gas clumps can migrate via dynamical friction to the centre of the galaxy and fuel rapid bulge formation \citep{Bournaud07,KrumDekel11}, while any mergers in the early universe can also stir the gaseous disk.  Given the importance of both processes in the life of disk galaxies, any indirect effort to constrain the maximum $\sigma_{ISM}$ at a given epoch is of vital interest.

A useful avenue in understanding feedback in the gas disk, and the influence of latent dynamical heating in the stellar populations, is to study the AVR in galaxies of a variety of masses.  Low mass dwarf galaxies are more susceptible to SNe feedback due to their lower gas densities and shallower potential wells, with some studies showing that even isolated dwarf galaxies may form their stars out of dynamically hot gaseous configurations at all times \citep{Kaufmann07,Wheeler16,Elbadry17}. Unlike higher mass rotationally supported disk galaxies where velocity dispersion is minor contributor to the orbital energy budget, any AVRs for the low mass dwarf spheroidals (dSphs; which can be dispersion supported), may indicate changes to the global gravitational potential, rather than a reflect the birth dynamics or impact of localized secular scattering process on the stars.  While interpretation of dSph AVRs will be more complex due to their environment near the MW, the dSphs offer a large lever arm to study mass dependent processes, and their vulnerability to feedback and tidally driven gravitational potential alterations makes study of their AVRs also a unique tool to understand their formation.

 Conversely higher mass gas disks will be thinner and more likely affected by disk instabilities, or turbulence from external gas accretion \citep{Dekel09}.  Dynamical heating of the stellar populations from spiral arms or giant molecular clouds (GMCs) should also be more prevalent in massive disk galaxies - both due to a mass dependent maximum GMC mass (c.f., \citealt{Kruijssen15}), and inability of spiral arms to form in thicker low density dwarf galaxy disks \citep{Streich16}.  For merging satellite galaxies, the mass spectrum of subhalos in $\Lambda$CDM is scale invariant, however the disk response of the host depends on the current disk structure and gas fraction \citep{Helmi12} - both of which may also lead to a mass dependence in this scattering process as well.

A natural next step is to exploit these mass scalings and extend the AVR analysis done in large spiral galaxies, to a sample of galaxies covering a range of masses.  Motivated by the well developed theoretical pictures for these processes, and the dual interpretation for the AVR shape (intrinsic from birth due to high $\sigma_{ISM}$ that stars are born out of, or generated over time by dynamical heating processes), in this work we analyze the AVR of eight Local Group galaxies.

Crucially, we seek to relax the canonical assumption in many AVR and disk heating studies, that stars are born from a cold disk with low velocity dispersion at all epochs.  The galaxies we consider span $\sim 5$ orders of magnitude in stellar mass, have AVRs based on high quality individual spectroscopic observations - and most importantly, all have complete star formation histories (SFHs) from colour magnitude diagram (CMD) analysis.  As we shall see this last criteria gives us an empirical backbone in order to construct physically motivated models to interpret the AVR, and in turn understand the dynamical evolution of low and high mass star forming galaxies.

\section{Data}
To study the velocity dispersion evolution in Local Group galaxies, we have compiled spectroscopic radial velocity data of individual stars from the literature, or in the case of M33 and the Milky Way (MW), stars and star clusters.  The spectroscopic observations are typically of evolved giant branch stars, which span ages of $2 \lesssim t \lesssim 13.0$ Gyrs.  In some cases (MW, M31, LMC, WLM) younger ($\leq 0.3$ Gyr) supergiant or main sequence turnoff stars are also incorporated in order to probe the recent dynamical evolution.

Table 1 shows the data sources for the dynamical analyses used in this paper.  We list the total number of stars with spectroscopic measurements, along with the approximate velocity uncertainty of the individual stellar spectra.  The spatial footprint of the spectroscopic samples typically is concentrated around $0.5-2 R_{e}$, or in the Solar neighbourhood for the MW sample.  For the MW we follow \cite{Martig14} and append to the \cite{Casagrande11} $\sigma$ values, the velocity dispersion of the bluest/youngest stars ($(B-V) \sim 0.0$) in \cite{Aumer09}, in order to allow for a more consistent differential comparison with the other galaxies where supergiant stars of comparable likely age are used. 

Analyzing the time evolution of the stellar velocity dispersion also requires the stellar spectra to have sufficiently high S/N for metallicities ([Fe/H]) to be derived. This allows us to break the age-metallicity degeneracy on the colour magnitude diagram (CMD) and compute relative ages for the stars - either via isochrone fitting (e.g., \citealt{Cole05}), or statistically via CMD analyses (e.g., \citealt{deBoer12}).  This procedure involves using a grid of stellar evolution models at fixed [Fe/H] and [$\alpha$/Fe] (in the case of the dSphs also constrained by independent spectroscopic measurements) but varying age, and interpolating the position of a star in the CMD to the closest isochrone.  The ages of the tracer populations are taken from the same literature sources as the kinematic data in Table 1.

Finally, we utilize observed star formation histories (SFHs) from analyses of CMDs (e.g., \citealt{Dolphin00}) to track how a galaxy's stellar mass grew over its lifetime.  An assumption in this is a choice of the IMF - the de Boer et al. sample here adopt a Kroupa IMF, others a Salpeter - however the choice is a smaller uncertainty than the error in the SFH model in most cases.  While a large number of SFHs exist for many galaxies in our sample, we tried to choose where possible, SFHs from a single study which covered a large field of view, and resolved the old main sequence turnoff (c.f., \cite{Gallart05}). As we wish to consider the evolution of global quantities such as $M_{*}$ and $M_{gas}$, it is necessary to homogenize the SFRs derived from CMD analyses of differing spatial footprints.  Therefore we must extrapolate the SFRs in the observed fields to that of the entire galaxy weighting by an exponentially declining radial SF density (using the gas disk scale length) in the case of M33 and M31.  The adopted areal correction factors are listed in Table 2.  As the increase in area in the outer regions will not change the total SFR significantly due to the radially concentrated SF disk, we could have also normalized the SFR of the observed fields to the stellar mass of the galaxy at present day from integrated photometry.  While our chosen method is an inherently uncertain estimate of the global SFR in cases where the field of view is small or where the disk growth history deviates strongly from inside-out, we stress that the final stellar mass growth histories using the adopted SFRs, are consistent within a factor of a few with the estimated stellar masses from integrated photometry or spectral energy distribution fitting, giving confidence to our estimates of the total SFR for any given galaxy.  

We note that this scaling by the present day area assumes minimal size growth for the galaxy at high redshift.  The effect of this assumption on the final gas masses is to underestimate the gas fraction by a factor of Area($z$)/Area($z=0$), which would be most important during a regime of rapid growth (e.g. $z>3$).  However given the uncertainty in early size evolution for galaxies and the possibly different physical conditions of the ISM (depletion time, stability criteria) which may counterbalance this effect, we simply urge caution in extending any models past a reasonable data calibrated regime (e.g., $t \gtrsim 12$ Gyrs).  In addition, this effect would only be significant if the galaxy had significant radial rearrangement (not mixing) of its stars over the course of a Hubble time.

For our adopted age bins and sample sizes, the random uncertainties on each velocity dispersion measurement will be minimal ($\leq 2$ km s$^{-1}$) for most of the galaxies.  Similarly, random uncertainties on the stellar ages computed via isochrone fitting are $\sim 50\%$, a conservative estimate typical of analyses in distant systems (e.g., \citealt{Leaman09,Hendricks14}).  However as with the velocities, systematic errors in the age dating of RGB stars dominate and can reach $1-3$ Gyrs due to uncertainties in [$\alpha$/Fe], differential reddening, and AGB star contamination (c.f., \citealt{Cole04, Cole05, Leaman13} for a detailed description of these effects).

\begin{table}
\caption{Galaxy Velocity Data}
\label{table:pars}
\begin{tabular}{@{}lccc}
\hline
Galaxy & $N_{stars}$ & $\langle\delta_{V}\rangle$ & Reference\\
\hline
\hline
Carina & 305 & 2 km s$^{-1}$ & \cite{Lemasle12}\\
\hline
Sculptor & 579 & 2 km s$^{-1}$ & \cite{Battaglia08b}\\
\hline
Fornax & 686 & 2 km s$^{-1}$ & \cite{Battaglia06}\\
\hline
WLM & 126 & 7 km s$^{-1}$ & \cite{Leaman12}\\
 & 19 & 10 km s$^{-1}$ & \cite{Bresolin06}\\
 \hline
LMC & 373 & 7 km s$^{-1}$ & \cite{Cole05}\\
 & 167 & 1 km s$^{-1}$ & \cite{Olsen07}\\
\hline
M33 & 48 & 8 km s$^{-1}$ & Beasley et al. (in prep)\\
 & 31 & 12 km s$^{-1}$ & \cite{Beasley15}\\
 \hline
 MW & 14139 & 1 km s$^{-1}$ & \cite{Casagrande11}\\
    & 579 & 2 km s$^{-1}$ & \cite{Hayes14}\\
 \hline
 M31 & 5800 & 10 km s$^{-1}$ & \cite{Dorman15}\\
\hline
\end{tabular}
\end{table}

Table 2 shows the sources for the SFHs, along with the areal factors used to extrapolate the observed SFRs to a global SFR for the galaxies.

\begin{table}
\caption{Galaxy SFH Data}
\label{table:pars}
\begin{tabular}{*{4}{p{1.7cm}}}
\hline
Galaxy & $m_{v} 50\%$ & $A_{obs}/A_{gal}$ & Reference\\
\hline
\hline
Carina & 23.5 & 1.0 & \cite{deBoer14}\\
\hline
Sculptor & 23.8 & 1.0 & \cite{deBoer12b}\\
\hline
Fornax & 23.9 & 0.46 & \cite{deBoer12}\\
\hline
WLM & $27.0$ & 0.11 & \cite{Dolphin00}\\
 \hline
LMC & $20.5$ & 1.0 & \cite{HZ09}\\
\hline
M33 & 27.2 & 0.033 & \cite{Barker07}\\
    & 27.0 & 0.033 & \cite{Barker07b}\\
 \hline
 MW & --- & 1.0 & \cite{Snaith15}\\
 \hline
 M31 & 27.4 & 0.5 & \cite{Lewis15}\\
     & $\sim 25$ & 0.003 & \cite{Williams03}\\
\hline
\end{tabular}
\end{table}

\begin{table}
\caption{Galaxy Integrated Properties}
\label{table:pars}
\begin{tabular}{@{}lllll}
\hline
Galaxy & $M_{*}$  [$M_{\odot}$] & $M_{HI}$  [$M_{\odot}$]& $M_{vir}$  [$M_{\odot}$] & Reference\\
\hline
\hline
Carina & $1\times 10^{6}$ & --- & $2\times 10^{9}$ & \cite{deBoer14}\\
\hline
Sculptor & $4\times10^{6}$ & --- & $3\times 10^{9}$ & \cite{deBoer12b}\\
    & & & & \cite{Amorisco14}\\
\hline
Fornax & $3\times 10^{7}$ & --- & $5\times 10^{9}$ & \cite{deBoer12}\\
       &                 &     &                 & \cite{Amorisco14}\\
\hline
WLM & $3\times 10^{7}$ & $1\times 10^{8}$ & $1\times 10^{10}$  & \cite{Jackson07a}\\
    &                  &                  &                   & \cite{Hunter11}\\
    &                  &                  &                    & \cite{Leaman12}\\
\hline
LMC & $3\times 10^{9}$ & $7 \times 10^{8}$ & $1\times 10^{11}$ & \cite{vdM02}\\
    & & & & \cite{Kim98}\\
    & & & & \cite{Gomez15}\\
\hline
M33 & $6\times 10^{9}$ & $1\times 10^{9}$ & $4\times 10^{11}$ & \cite{Corbelli03}\\
    & & & & \cite{Seigar11}\\
\hline
MW & $5\times 10^{10}$ & $8\times 10^{9}$ & $1\times 10^{12}$ & \cite{McMillan11}\\
   & & & & \cite{Kalberla07}\\
   & & & & \cite{Budenbender15}\\
\hline
M31 & $9\times 10^{10}$ & $7\times 10^{9}$ & $2\times 10^{12}$ & \cite{Fardal13}\\
\hline
\end{tabular}
\end{table}

\begin{figure*}
\begin{center}
\ifpdf
\includegraphics[width=0.78\textwidth]{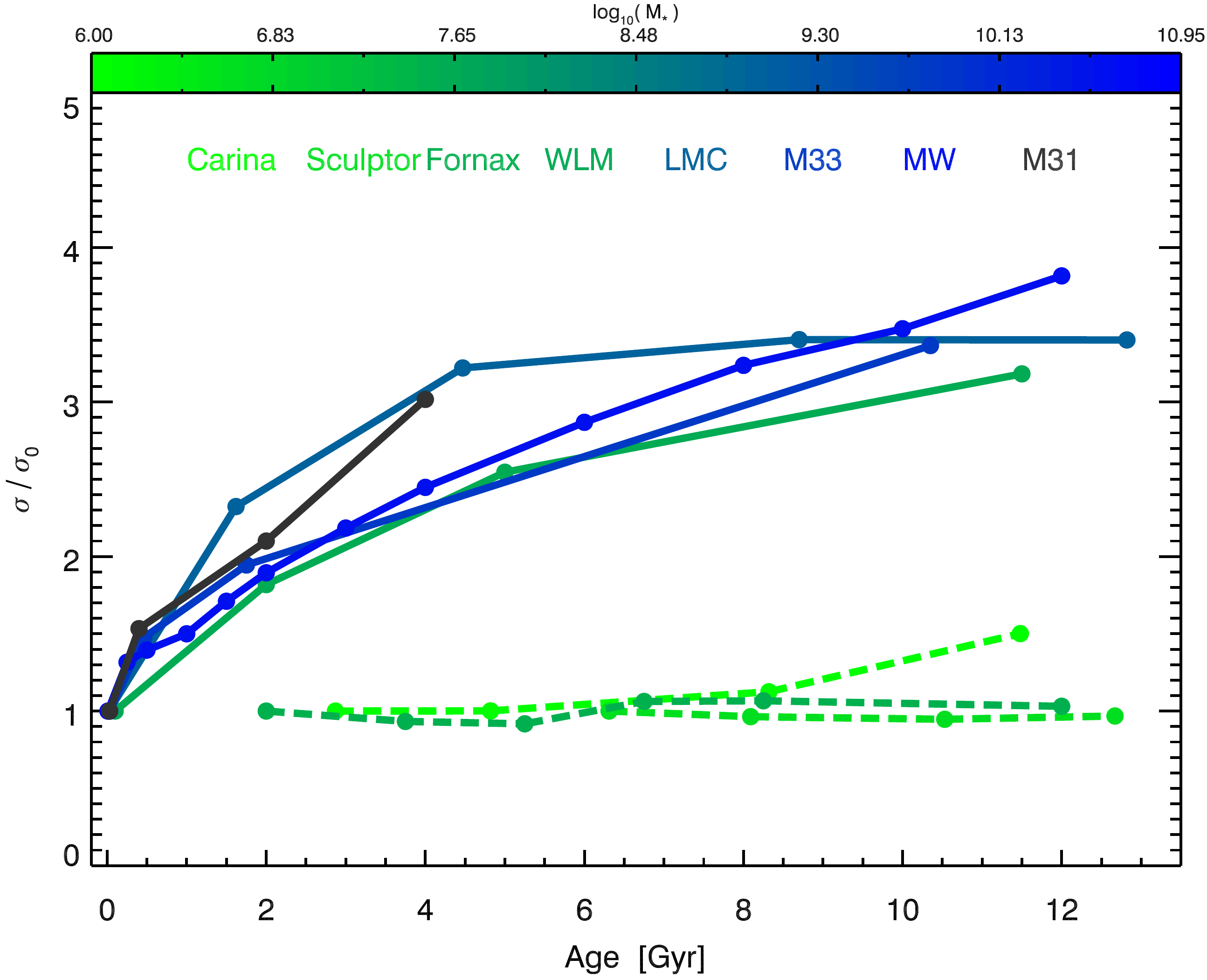}
\else
\includegraphics[width=0.78\textwidth]{normdispf.ps}
\fi
\caption{Normalized $\sigma(t)$ curves for the galaxies considered in this work.  The age-velocity curves have been normalized to the stellar velocity dispersion of the youngest stars for each system.  Dwarf spheroidals are shown as dashed lines, and show a noticeable lack of evolution.}
\label{fig:normdisp}
\end{center}
\end{figure*}

\section{Analysis of the Stellar Kinematics and Ages}
For a given galaxy, the sample of spectroscopic stars are first divided into age bins.  Given the large systematic errors on both velocity dispersion and age (see Section 2), we have checked that our results are insensitive to the choice of binning.  In the star forming galaxies with significant rotation, the bulk velocity field is removed so that the stars are in a rest frame with respect to the galaxy's systemic velocity, and spurious enhancements in dispersion are not introduced from considering stars on approaching and receding sides of the velocity field.  Within each age bin, the stellar velocity dispersion and its confidence interval are calculated in a maximum likelihood fashion following the methodology in \cite{Walker06}.

As the LOS velocities in each sample differ in their projection, we must produce a homogeneous velocity dispersion measure in order to compare between galaxies at different inclinations.  If the velocity ellipsoid of the galaxy is anisotropic, the projected dispersion has a dependence on the amount of velocity anisotropy and viewing angle.

We compute the total velocity dispersion $\left(\sigma = \sqrt{\sigma_{\phi}^{2} + \sigma_{R}^{2} + \sigma_{z}^{2}}\right)$ as:
\begin{equation}
\sigma = \left(\frac{\sigma_{LOS}^{2}} {1 - \delta(\rm{cos(}i\rm{)}^{2})}\right)^{1/2},
\end{equation}
where $\delta$ is a joint anisotropy parameter describing the shape of the velocity ellipsoid and is 0 for an isotropic system.
\begin{equation}
\delta = \frac{2\gamma - \beta}{2 - \gamma}
\end{equation}
With $\beta = 1 - (\sigma_{\phi}/\sigma_{R})^{2}$ and $\gamma = 1 - (\sigma_{z}/\sigma_{R})^{2}$.

The bulk of the error budget for the $\sigma$ values comes from the \textit{systematic} uncertainty due to the poorly constrained inclination and velocity anisotropy in the systems we consider.  In some galaxies there exist independent estimates of the velocity tensor from proper motion (the MW \citealt{Jenkins90, Hattori11}), or coupled gas/stellar kinematic analysis  (WLM; \citealt{Leaman12}, c.f., \citealt{Silchenko11}).  In other cases such as the dSphs or LMC, the inclination or anisotropy are difficult to constrain.  

For M31 and M33, we could assume that the better known anisotropy parameters of the MW disk would also apply to these spiral disk galaxies, given that studies of external galaxies have found similar values (e.g., \citealt{Shapiro03}).  However we chose to also be agnostic with respect to these galaxies, when direct measurements were not available, and we adopt a range that roughly spans the limit seen in measurements of the stellar velocity ellipsoids in a recent literature compilation (Pinna et al. 2017).  While more stringent constraints will continue to become available in the future, reduction of the anisotropy limits to for example $-0.3 \leq \beta \leq 0.5$ only changes the uncertainties on the marginalized velocity dispersion by $\sim 5 - 15\%$ depending on the inclination and value of $\delta$.  Table 4 shows the adopted ranges of inclination and anisotropy parameters $\beta, \gamma$ used in computing the homogenized total dispersion $\sigma$ using Equations 1 and 2.

The final error bars on the dispersion reflects the maximum and minimum $\sigma$ value given the range of anisotropy and inclination parameters used.  This is added in quadrature to the statistical random uncertainty.  The average of the minimum and maximum is is taken as the final estimate of the deprojected total velocity dispersion value.
 
 \begin{table}
\caption{Anisotropy and Inclination Ranges}
\label{table:pars}
\begin{tabular}{@{}lcccl}
\hline
Galaxy & $\beta$ & $\gamma$ & $i$ & Reference\\
\hline
\hline
Carina & [-0.7,0.5] & [-0.7,0.5] & [1,80] & ---\\
\hline
Sculptor & [-0.7,0.5] & [-0.7,0.5] & [1,80] & ---\\
\hline
Fornax & [-0.7,0.5] & [-0.7,0.5] & [1,80] & ---\\
\hline
WLM & [-0.7, 0.3] & [0.0,0.4] & [68,88] & \cite{Leaman12}\\
\hline
LMC & [-0.7,0.5] & [-0.7,0.5] & [21,40] & \cite{vdM02}\\
\hline
M33 & [-0.7,0.5] & [-0.7,0.5] & [51,61] & \cite{Zaritsky89}\\
\hline
MW & [0.3,0.7] & [0.3,0.8] & [50,90] & \cite{Hayes14}\\
 & & & & \cite{Jenkins90}\\
 & & & & \cite{Hattori11}\\
 & & & & \cite{Shapiro03}\\
 \hline
M31 & [-0.7,0.5] & [-0.7,0.5] & [73,80] & \cite{Dorman15}\\
\hline
\end{tabular}
\end{table}

\section{Observational Evidence for Velocity Dispersion Evolution}
In Figure \ref{fig:normdisp} we show the age-velocity dispersion relations (AVRs; $\sigma(t)$) of all the galaxies normalized to their velocity dispersion of the youngest stars.  This figure gives a visual impression of the general self-similar shape for the $\sigma(t)$ curves amoung the gas rich star forming galaxies. These galaxies tend to consistently show strong $\sigma_{v}$ evolution in the first few Gyrs, eventually all arriving at a value of $\sigma/\sigma_{0} \sim 3$ for the oldest stars.  Notably the dSphs show approximately flat dispersion profiles at all ages, which will be discussed in Section 7.5.

The velocity dispersion evolution implied by the $\sigma(t)$ curves in Figure \ref{fig:normdisp} may be interpreted as due to a latent dynamical heating of stars that were born out of an ISM that was cold at all times (e.g., $\sigma_{ISM}(t) \leq 10$ km s$^{-1}$).  Alternatively the stars which formed at high redshift may have been born in much more turbulent gas than is observed in galaxies at present day, and therefore the oldest stars would exhibit higher velocity dispersions than the youngest stars.   In this interpretation some process such as SNe feedback, radiation pressure, dynamical friction driven gas flows or magnetic or gravitational instabilities in the gas disk may generate the higher turbulent velocity dispersions at early times (c.f., \citealt{Bournaud15}).  In the following sections we explore a simple, observationally motivated model for the evolution of the gas velocity dispersion with redshift, and discuss how and when the stars decouple from this velocity field after their birth.

\section{Model Description}
There is statistically significant evidence (e.g., \citealt{Wisnioski15} and references therein, hereafter W15) that galaxies have larger gas velocity dispersions at high redshift - with average values of ionized gas dispersions dropping by a factor of $\sim 2$ from $\sigma_{H\alpha} \sim 50$  at $z \sim 2$ to present\footnote{The typical random uncertainties on velocity dispersion for local $z=0$ galaxies in these studies are $\delta_{\sigma} \pm 2$ km s$^{-1}$ and $\delta_{\sigma}/\sigma \sim 25\%$ for the high redshift galaxies}.  Older stars may therefore be born with intrinsically higher velocity dispersions if the large scale gas kinematics out of which they formed, were more disordered in the early universe. 

Simultaneously, simulations (e.g., \citealt{Grand16,Aumer16} have shown, that time dependent scattering mechanisms can shape the AVRs of stars after they are born (``heating histories''), and these scattering processes themselves likely vary in efficiency over time.

Therefore in order to understand the physical mechanism(s) producing the $\sigma(t)$ curves in the galaxies, we attempt to model the intrinsic dynamical \emph{cooling} in the velocity dispersion of the ISM with time as well as scattering effects which should alter the orbits of the coldest population of stars. This composite model will provide insight into velocity dispersion of stars at birth and subsequent dynamical heating in galaxies in a range of masses.

\subsection{$\sigma_{ISM}$ Evolution}
\begin{figure}
\begin{center}
\ifpdf
\includegraphics[width=0.47\textwidth]{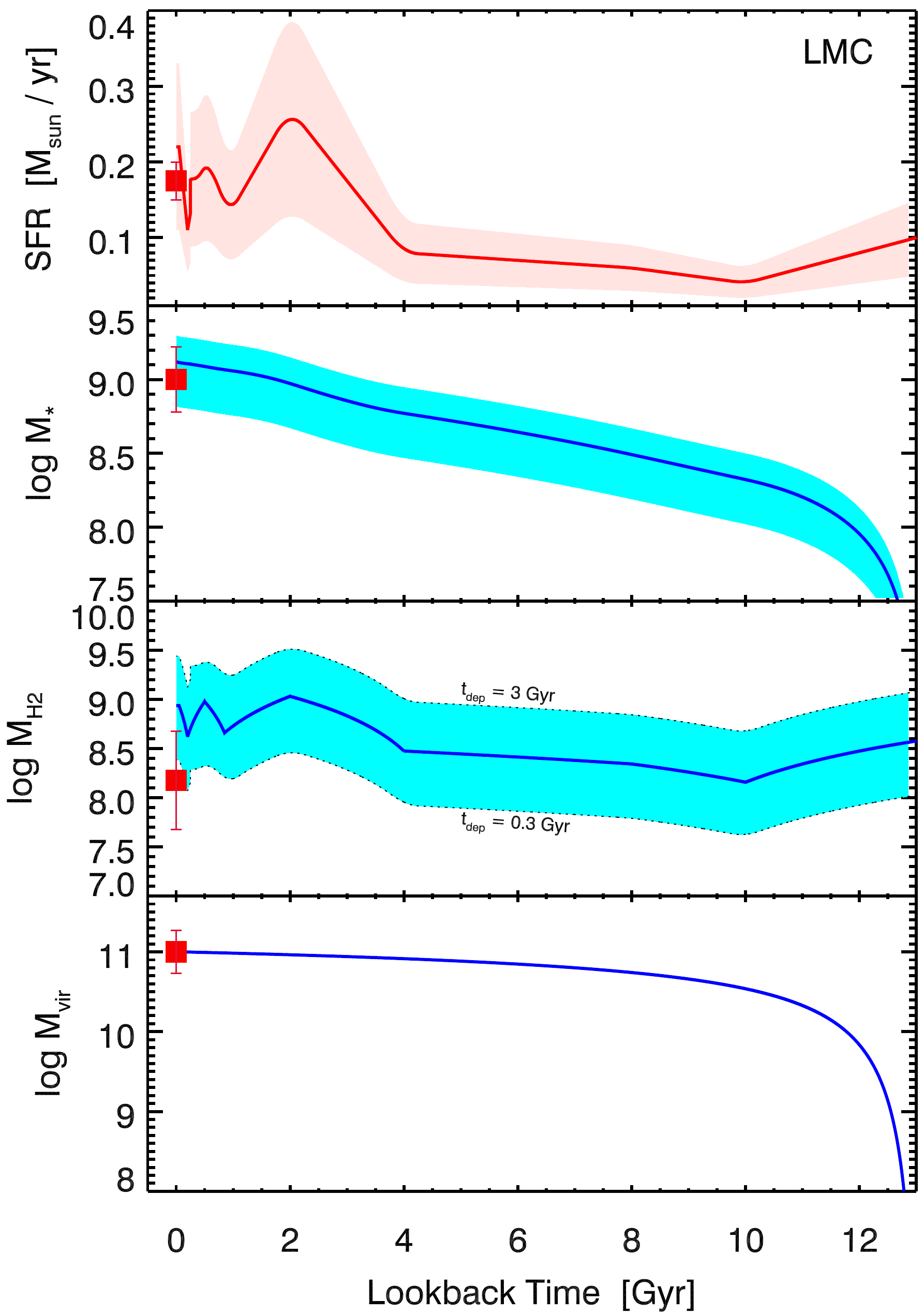}
\else
\includegraphics[width=0.47\textwidth]{oneheatevof.ps}
\fi
\caption{Observed SFH and derived evolutionary model tracks for an example galaxy (the LMC).  In particular the current gas mass is a strong constraint on the chosen star formation efficiency (parameterized as a gas depletion time here) within the models, providing a check that the evolutionary models recover the present day observed values (shown as red points at $t=0$).}
\label{fig:oneheat}
\end{center}
\end{figure}

We draw upon the recent analysis from the KMOS$^{\rm{3D}}$ team, who have produced empirical relations of the ISM velocity dispersion versus lookback time consistent with galaxies in several spectroscopic surveys (W15).  The galaxies observed extend out to $z \sim 4$ and represent the most detailed look at the environment in which stars are being born for galaxies of mass $10^{9} \leq M_{*} \leq 10^{12}$.  W15 showed that the observed gaseous kinematic data show a decrease in velocity dispersion that is consistent with expectations based on gravitational instability criteria for star forming disks with an evolving gas fraction (see also, \cite{Kassin12,Kassin14}).

Figure 10 and Equation 8 of W15 show that galaxies of different rotational velocities occupy roughly parallel sequences in a plot of $\sigma_{ISM}$ versus redshift.  However this figure is an instantaneous snapshot of the ISM conditions for galaxies of a given mass or circular velocity at those epochs.  We would expect that galaxies should evolve through a different path in this parameter space as they change in baryonic mass and circular velocity with time.  The two key parameters in the model of W15 which control the ISM velocity dispersion are the gas fraction, and the rotation velocity of the galaxy:

\begin{equation}
\sigma_{ISM}(z) = \frac{Q}{a}V_{rot}(z)f_{gas}(z),
\end{equation}

where $Q$ is the Toomre stability parameter and $a$ a factor which depends on the shape of the velocity field.  These may plausibly vary over the range $0.67\leq Q \leq 2.0$ and $1 \leq  a \leq \sqrt{4}$, however we adopt $Q=1, a=\sqrt{2}$ for the baseline model.

If not directly measured, the gas fraction can be expressed in terms of the specific star formation rate (sSFR), and depletion time ($t_{dep}$).  W15 used empirical scaling relations between $M_{*}-sSFR$ and $M_{*}-t_{dep}$ to indirectly compute $f_{gas}$ for their objects, and show that the $\sigma_{ISM}$ evolution was consistent with disk instabilities driving the turbulence in galaxies at different epochs.

Rather than use similar empirical scaling relations (which may not be valid at $z=0$ or below the mass limit of the observational samples they were defined by), we will infer the gas fraction directly using the observed SFHs from the CMD analysis of our Local Group galaxies.  This allows for a self-consistent analysis that uses explicit observational data to construct a disk instability model, and doesn't require that the galaxy evolve along a static star forming main sequence.

Integration of the CMD based SFH provides the evolution of the galaxies' stellar mass with time, $M_{*}(t)$ (modulo a remnant fraction).  To track the gas mass, we can ask what amount of gas would be needed in the galaxy to allow it to form the requisite number of stars in the next timestep.

The gas mass in this formulation of course depends on the efficiency with which gas is converted into stars.  There are various parameterizations which link the surface density of gas to the surface density of SFR, with a general power law dependence of $\Sigma_{SFR} \simeq \epsilon_{SF} \Sigma_{gas}^{n}$ suggested by many studies \citep{Leroy08,Bigiel08}, though the super or sub-linearity of the power law exponent in the relation, as well as the best gas tracer (HI, H$_{2}$, CII, HCN) is still an active area of discussion \citep{Glover15}.  Following \cite{Bigiel08}, we adopt $n=1.0$ for the molecular gas exponent.  The proportionality constant represents a conversion efficiency, with an equally broad set of physical interpretations, but is commonly expressed as an inverse of the gas depletion time, $\epsilon_{SF} = t_{dep}^{-1}$.

Together with our observed SFRs as a function of time from the CMD analysis, we can invert this Schmidt-Kennicutt relation to obtain the molecular gas mass - provided we specify a depletion time.  As stars form out of the cold molecular phase, we specify the molecular gas mass directly and avoid having to estimate an molecular fraction within the model.

\begin{equation}
\Sigma_{H2} = t_{dep,H2}(\Sigma_{SFR})^{1/n}
\end{equation}

To set the range of plausible depletion times we consider the present day molecular gas masses of the SF galaxies in our sample, together with estimates of the depletion time from the literature studies of these and other nearby galaxies \citep{Leroy08}.  We also compute the depletion time directly using the observed SFRs and molecular gas masses.  Based on these we find a range of molecular gas depletion times from $0.3 \leq t_{dep,H2} \leq 3$ Gyr, which span the properties of our galaxies, while providing close agreement with studies of galaxies in the Local volume and at high redshift \citep{Saintonge11}.  The gas fraction in Equation 3 can now be computed self consistently from the CMD based SFHs, with the stellar mass and molecular gas mass explicitly setting $f_{gas} = M_{H2}/(M_{H2} + M_{*})$.

To estimate the redshift evolution in the galaxy's rotation velocity, we utilize semi-analytic halo growth models of \cite{Dutton09} and \cite{Dutton11} (hereafter D11) which describe on average, how the disk and halo properties of a galaxy of a given virial mass may evolve in redshift for a specified CDM cosmology\footnote{We adopt $\Omega_{m} = 0.3, \Omega_{\Lambda} = 0.7, H_{0} = 72$ \citep{Freedman01}}. These halo properties form a backbone for computing the rotation velocity at a given redshift and radius within the galaxy.  We calibrate and check that the growth models are appropriate, by requiring that they reproduce the present day observed and derived properties of the galaxies (stellar mass $M_{*}$, HI gas mass $M_{gas}$, and virial mass $M_{vir}$) (see Figure \ref{fig:oneheat}).  The adopted values and references which we calibrate the halo growth models to are shown in Table 3. 

The evolution of a single galaxy's \emph{molecular} gas velocity dispersion in the $\sigma_{ISM}$ redshift plane can now be predicted from Equation 3.  The strength of this evolutionary model for the gas velocity dispersion, is that it is physically motivated based on a disk instability argument, and for our sample of galaxies is computed using the observed SFHs.  We note that we find good agreement in the predicted $\sigma_{ISM}$ if we also use our SFH based $M_{*}(t)$ values in the empirical scaling relations for $f_{gas}$ as done in W15. The models presented are for a radius of $\sim 1 R_{e}$, which is comparable to where most of the kinematic tracers reside in each of the galaxies, however the models can be fine tuned as a function of radius if desired, as is done for comparison to the MW sample which is restricted to small volume.

The ISM model formally can only describe evolution of the velocity dispersion due to changing gas fractions within a growing potential, when the velocity dispersion component is sub-dominant in the orbital energy budget - for these models the validity is $V/\sigma \gtrsim 0.5$.  Application of these models to the dSphs then implicitly assumes that they evolved most of their life as 'disky' dwarfs, comparable to the dIrrs, but at some point exhausted their gas supply.  While classically described as spherical, dispersion support systems, newer observations are finding remarkable dynamical complexity \citep{Amorisco12,Kacharov17}, comparable angular momenta and global chemistry to dIrrs \citep{Wheeler16,Leaman13,Kirby13}, which when taken with simulations suggesting common formation and evolution for dIrrs and dSphs, justify application of our model to these systems.  The caveats mention above should be kept in mind however, and will be discussed in Section 7.4 with respect to the sensitivity of these systems to tertiary processes such as stellar feedback and tidal heating.

\subsection{Scattering of Stars by Disk Overdensities}
A fundamental assumption in interpreting the AVR as due to dynamical cooling of the ISM, is that newly formed stars are born with the velocity dispersion of the gas at that epoch.  While this may be true, we need to consider the fact that stars born with extremely cold gas dispersions, of order $1-5$ km s$^{-1}$, will not remain at that velocity dispersion in equilibrium (c.f., \citealt{Ambartsumian55}). 

Since the pioneering work of \cite{SS51} it has been argued that potential fluctuations could exist in the form of a population of (giant) molecular clouds in the disk.  Alternatively spiral arms \citep{Sellwood02} could also produce scattering and dynamical heating of disk stars (though see \citealt{Minchev12}). Such dynamical scattering will be most effective for populations born very cold - as their intrinsic dispersion can be altered by even small potential overdensities. Therefore stars in such thin, dynamically cold configurations will be extremely sensitive to even minimal overdensities (higher mass stars, binaries, associations, molecular clouds) and rapidly scatter.

We note that additional energy sources may provide feedback in the ISM to prevent $\sigma_{ISM}$ from becoming as low as the pure gravitational disk instability model predicts (e.g., \citealt{Tamburro09,Stilp13}), and will be discussed in Section 7.2.2.  Here we simply quantify the limiting case of scattering from a cold disk. As we shall see the velocity dependence of the stellar diffusion ensures that a robust prediction for the equilibrium state may be computed, regardless of whether the additional SNe feedback produces a slightly hotter disk.

In either case the thinnest, coldest stellar populations will be affected most strongly, and for the longest period of time, as any mid-plane disk scattering process becomes somewhat self limiting \citep{Wielen77}. As a star gets scattered increasingly more by a perturbing population, it spends less and less time in the plane of the disk and therefore the interaction time decreases, as does the subsequent dynamical heating.

Such a mid-plane scattering mechanism is therefore time dependent \citep{Aumer16}, and also has an inverse velocity dependence arising from the scaling of impact parameters in two-body interactions.  From first principles it can be shown that for a population of stars on circular disk-like orbits encountering a population of perturbers, the increase in velocity dispersion scales as \citep{SS51,Wielen77}:

\begin{equation}
\sigma = (\sigma_{ISM}^{3} + \frac{3}{2}\gamma t)^{1/3},
\end{equation}
where $\sigma_{ISM}$ is the initial birth velocity of the stars, and $\gamma$ the velocity dependent diffusion coefficient (with units of (km/s)$^{3}$ yr$^{-1}$), which is inversely proportional to the  disk dynamical relaxation time ($t_{rel}$) in the presence of a scattering population \cite{SS53}:

In order to describe the efficiency of the stellar scattering, we need to estimate the parameter $\gamma$ (or equivalently the quantities to solve for $t_{rel}$).  Past studies \citep{Wielen77, Sellwood02}, have typically fit for this parameter, by requiring that all stars be born with a floor ($\sigma_{ISM} \sim 5-10$ km s$^{-1}$) and that dynamical scattering alone provide the necessary increase in dispersion to match the present day observed AVR.  This assumes that there is no evolution in $\sigma_{ISM}$, an assumption which is increasingly inconsistent with observations of high-redshift galaxies and which we wish to relax, as the SFH based ISM cooling models we utilize here provide an independent measure of the dispersion with which the stars are born at, $\sigma_{ISM}(t)$.  

To independently calibrate $\gamma$ we utilize the typical scale heights of disk galaxies, and apply a disk equilibrium argument ($h = \sigma_{*,thin}^{2}/(\pi G \Sigma)$) to estimate the equilibrium stellar velocity dispersion corresponding to the thinnest scale height observed for a given mass galaxy.  The velocity dispersion that our stars are born at in the ISM cooling model $\sigma_{ISM}$ must eventually be increased by a scattering process to this equilibrium floor $\sigma_{*,thin}$ in approximately a dynamical time.  The diffusion coefficient can then be calculated directly for this limiting case as $\gamma = ((\sigma_{*,thin}-\sigma_{ISM})^{3}/t_{dyn,disk})$.

We use the data of \cite{Kregel02} who examined the ratio of disk scale length to scale height as a function of galaxy stellar mass ($8.3 \leq {\rm log_{10}}(M_{*}) \leq 11.5$) for a sample of galaxies.  While this ratio depends on the particular stellar population which is traced by that photometric band (e.g., \cite{Bovy12}), we take the upper envelope of the data to constrain what the thinnest galaxy at a given mass is.  The general trends in this study are also consistent with Local studies of disk thickness by \cite{YD06,Roychowdhury10,SJ10}.   The dynamical time of the disk is computed by creating rotation curves of the gaseous and stellar components using the exponential surface density distribution parameters specified by the growth models of D11 and our observed and inferred stellar and gaseous masses.

This exercise provides diffusion coefficients that are in the range $6\times10^{-8} \leq \gamma \leq 1\times10^{-5}$.  These estimates are in good agreement with an alternate method (see $\S 7.1$ Figure \ref{fig:sigratf}, Appendix B), whereby we compute the equilibrium timescale for a scattered population directly, following \cite{SS53}:
\begin{equation}
t_{rel} = \frac{\sigma_{0}^{3}}{\gamma} = \frac{4\sigma_{0}^{3}}{3\pi^{3/2}G^{2}n_{GMC}M_{GMC}^{2}ln(C)F(B/\omega)}.
\end{equation}
Here the logarithm takes the value of $\simeq 3$ and $F(B/\omega) = 0.98$ is a function of the local Oort constant and angular velocity, but which is invariant across our sample provided all galaxy rotation fields result from exponential disks evaluated at comparable fractions of the disk scale length. The the number density of clumps, $n_{GMC}$ and clump mass in Equation 6 are estimated directly from our ISM model assuming $M_{GMC}$ equals the largest unstable Jeans mass at any timestep in the ISM, and the cloud-cloud spacing is comparable to the Jeans length of the same largest unstable mode (see also Appendix B).

Our independent calibration for the scattering efficiency is based on the limiting case where the $\sigma_{ISM}$ is only due to the gravitational instabilities.  However, as mentioned at the start of this section, roughly the same (within $\sim 10$\%) equilibrium velocity dispersion for the stars is achieved if we add in a secondary source of dispersion (see Section 7.2.2), primarily due to the velocity dependence of the stellar diffusion.  Therefore we consider the equilibrium diffusion model applicable whether or not we considered additional sources of turbulence in the ISM.  In addition the procedure is only relevant for the low $\sigma_{ISM}$ dispersions predicted in some of the dwarf galaxies in our sample, and irrelevant for the higher mass galaxies.  Relaxing the assumption of stars being born from a constant dispersion floor, and instead letting them be born as indicated by the SFH based ISM model in a physically motivated fashion, yields diffusion coefficients which are consistent and in better agreement with independent semi-empirical GMC population models for scattering.

\begin{figure*}
\centering
\subfigure{\label{fig:oneheatcar}\includegraphics[width=62mm]{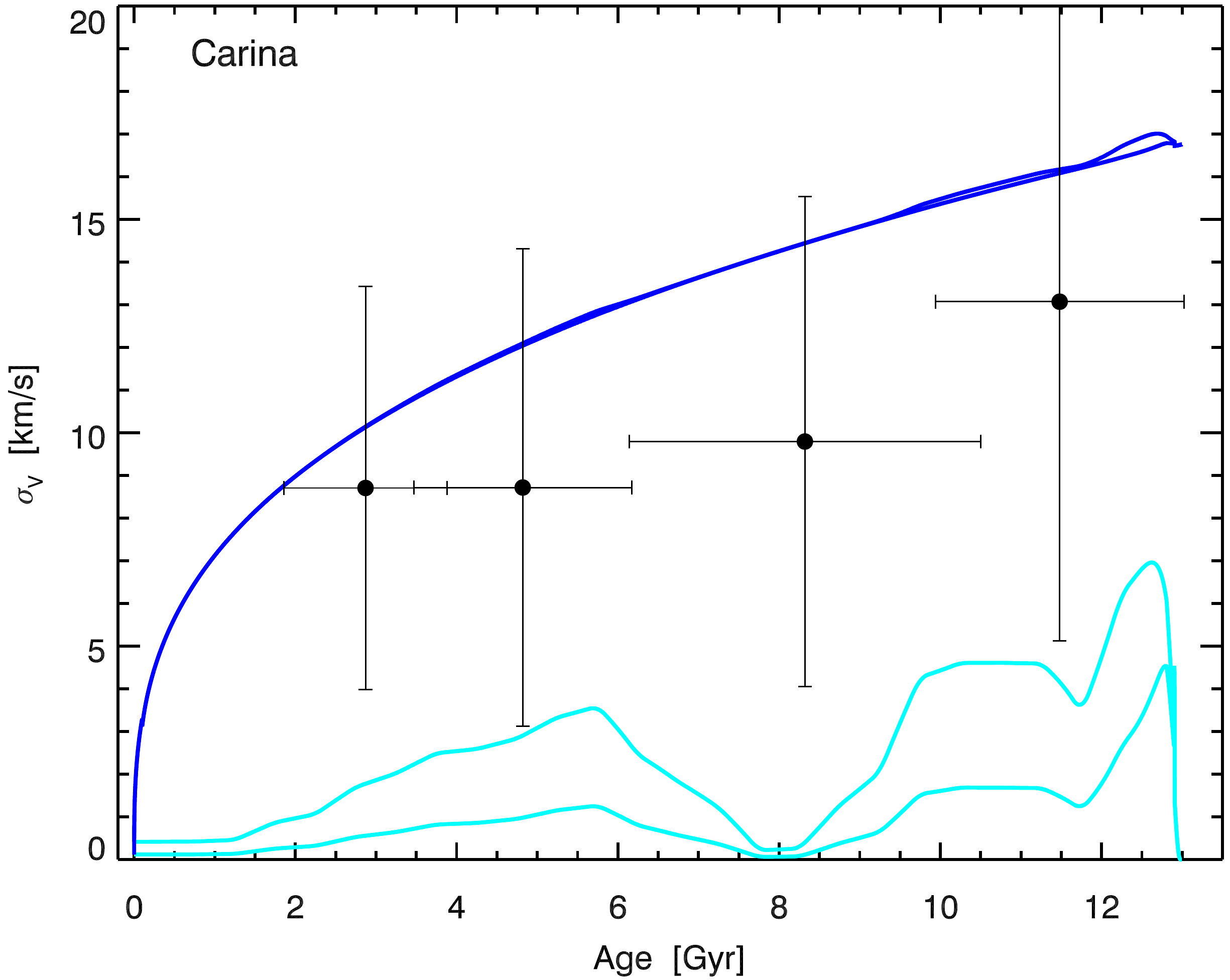}}
\subfigure{\label{fig:oneheatscl}\includegraphics[width=62mm]{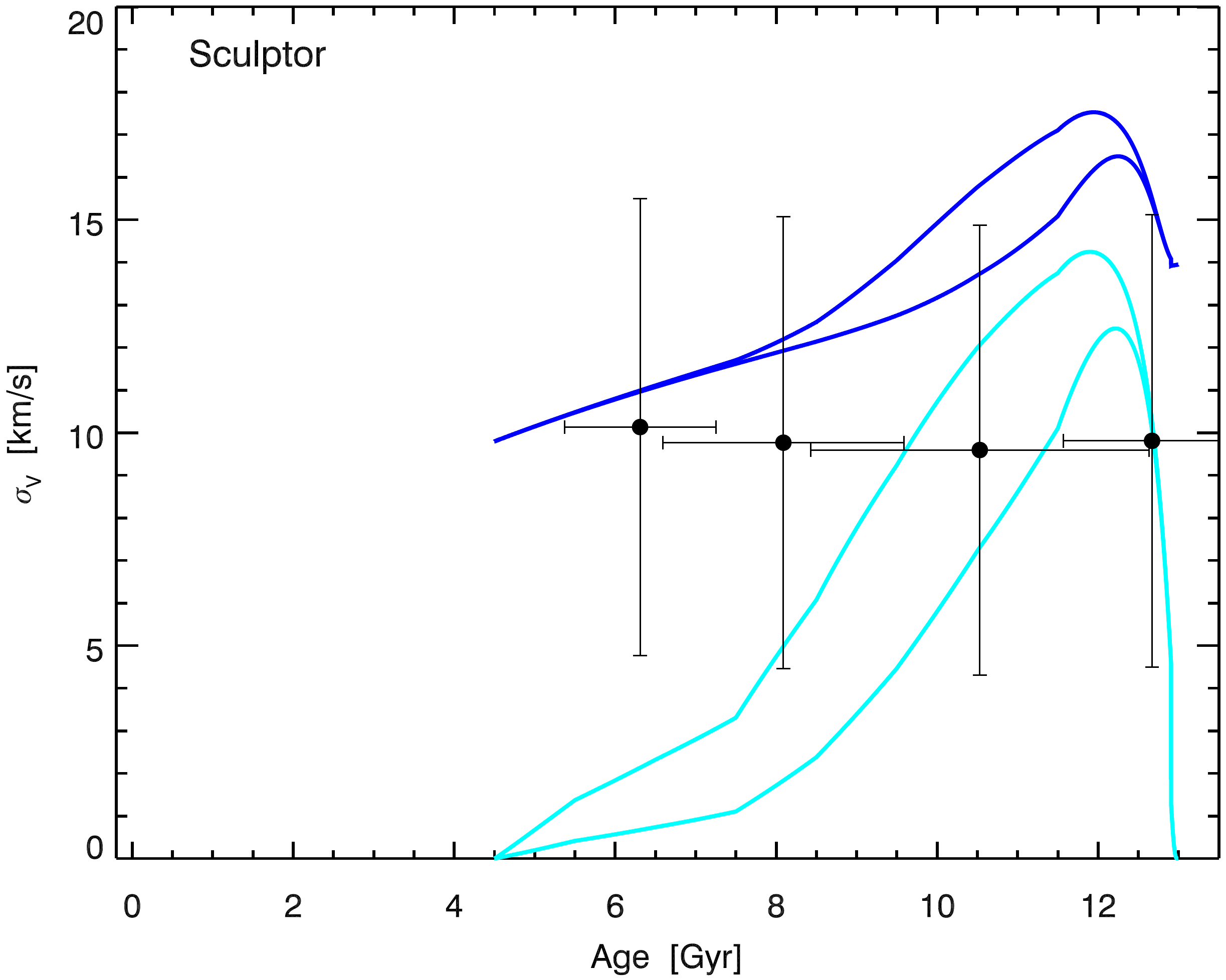}}\\
\subfigure{\label{fig:oneheatfor}\includegraphics[width=62mm]{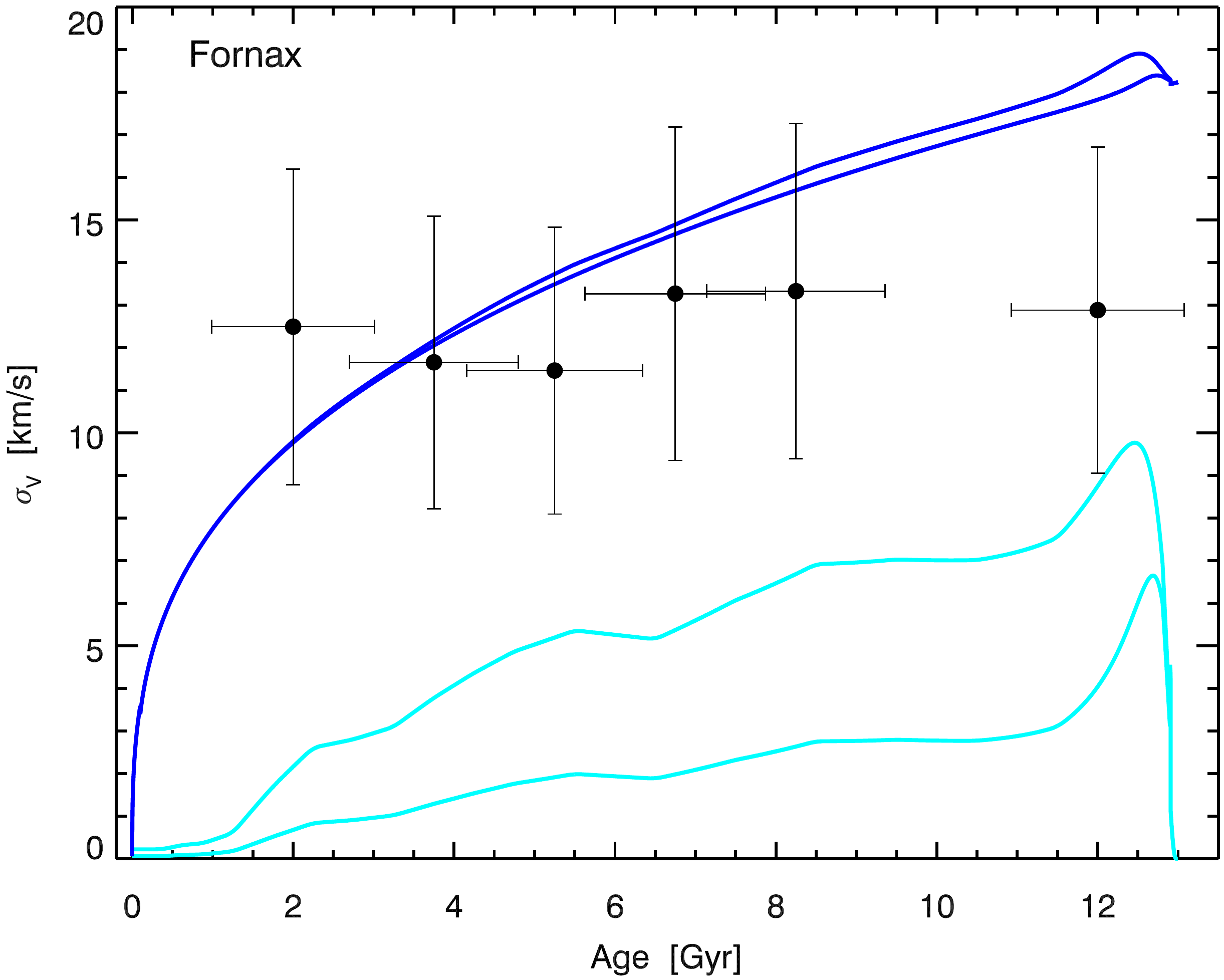}}
\subfigure{\label{fig:oneheatwlm}\includegraphics[width=62mm]{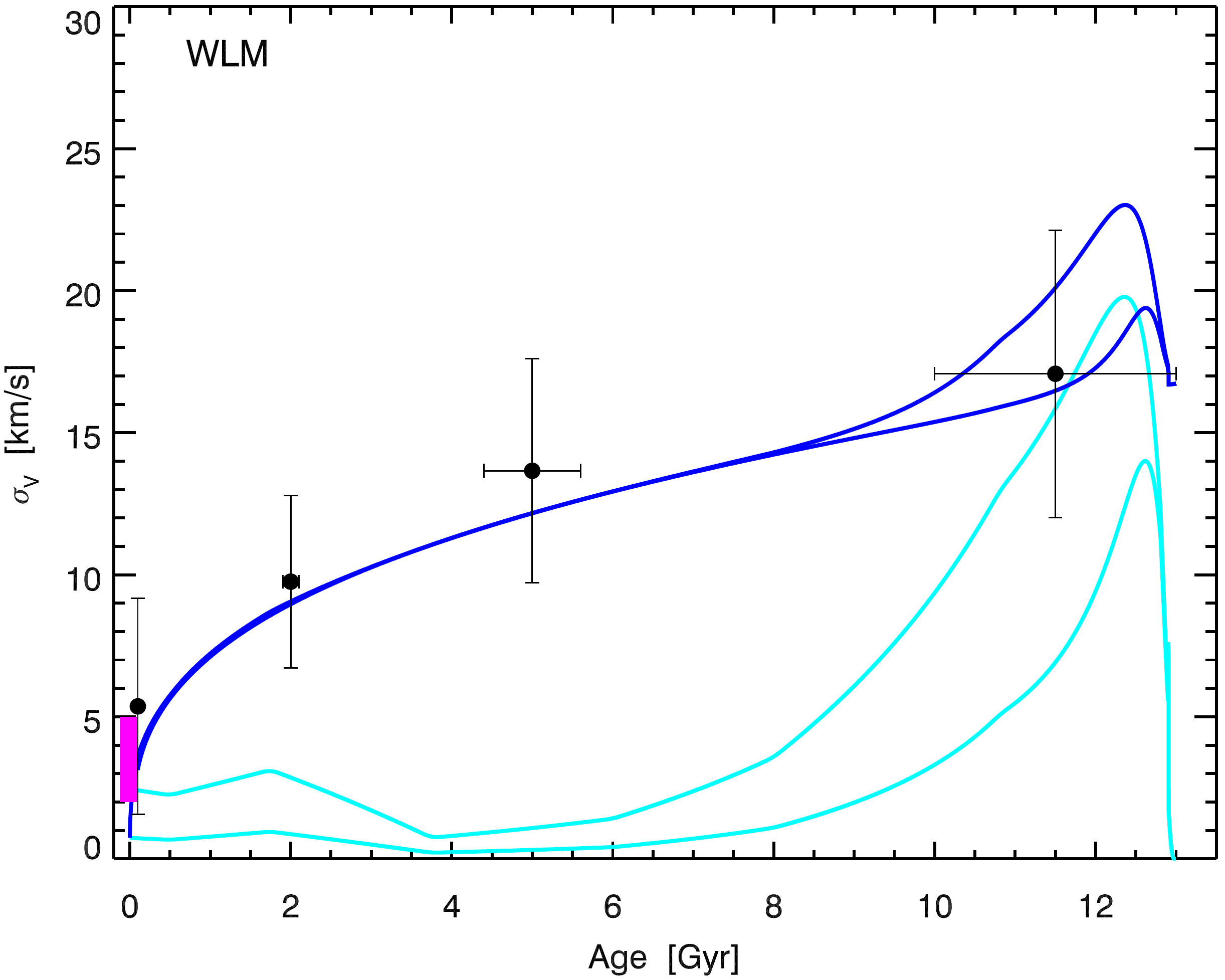}}\\
\subfigure{\label{fig:oneheatlmc}\includegraphics[width=62mm]{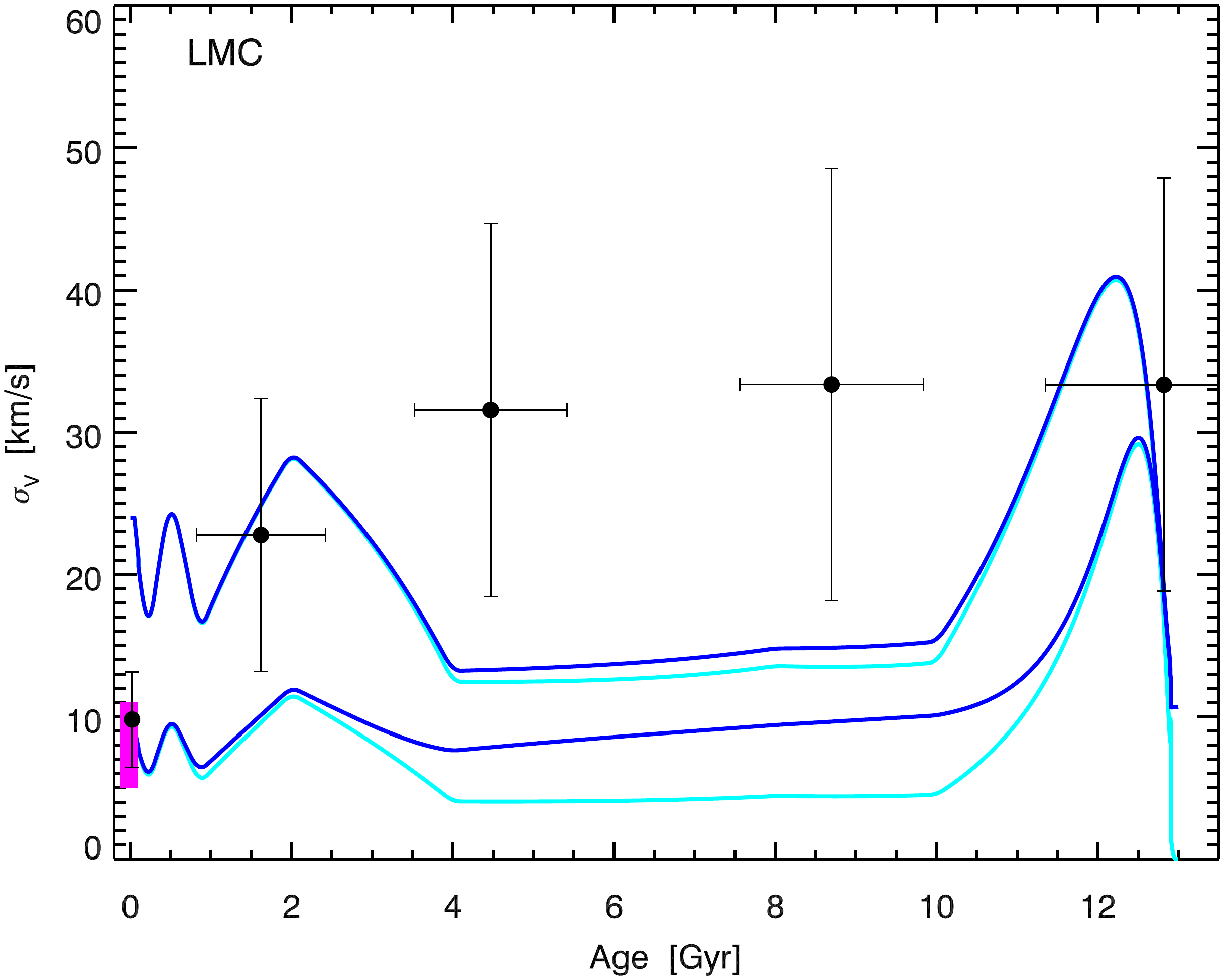}}
\subfigure{\label{fig:oneheatm33}\includegraphics[width=62mm]{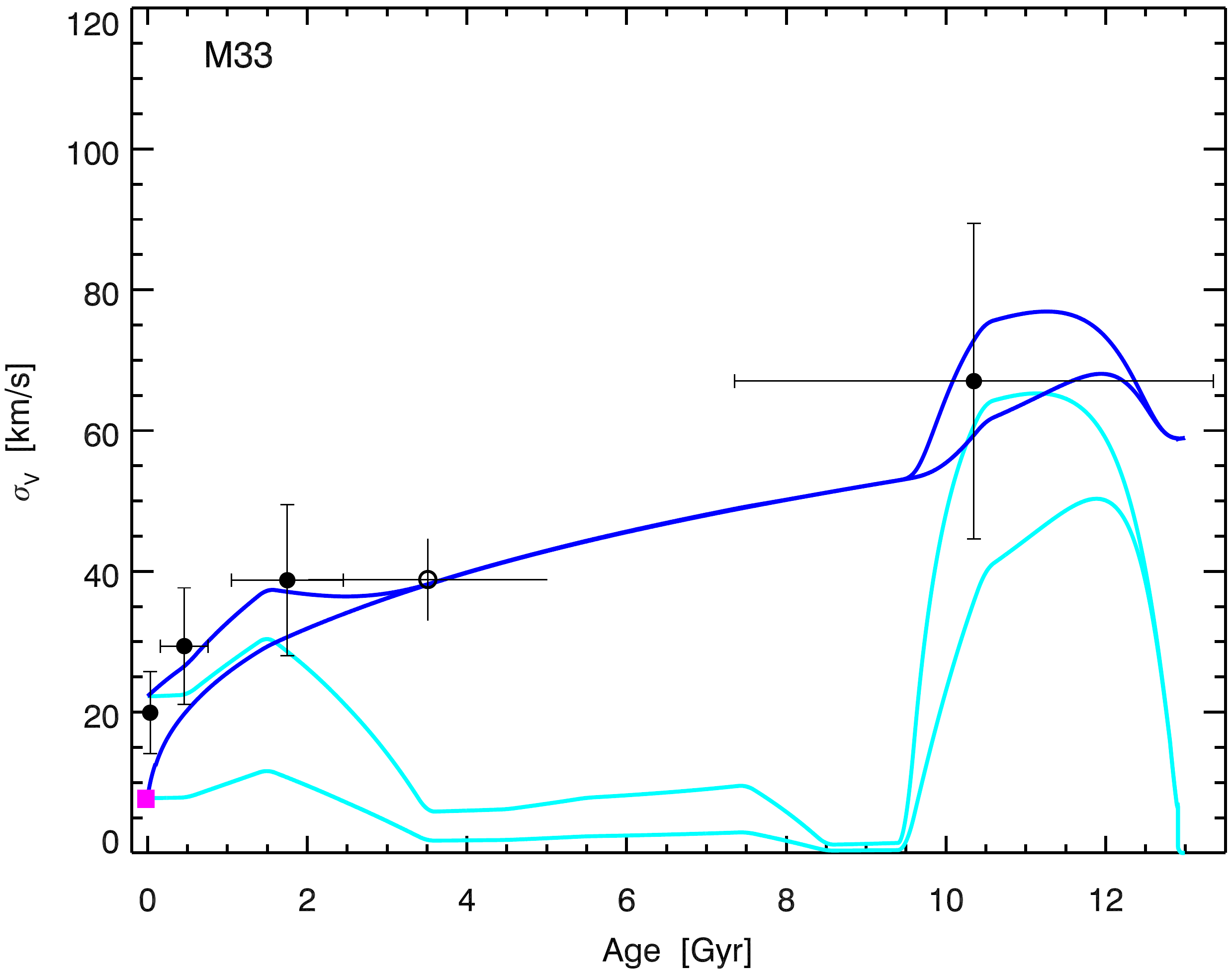}}\\
\subfigure{\label{fig:oneheatmw}\includegraphics[width=62mm]{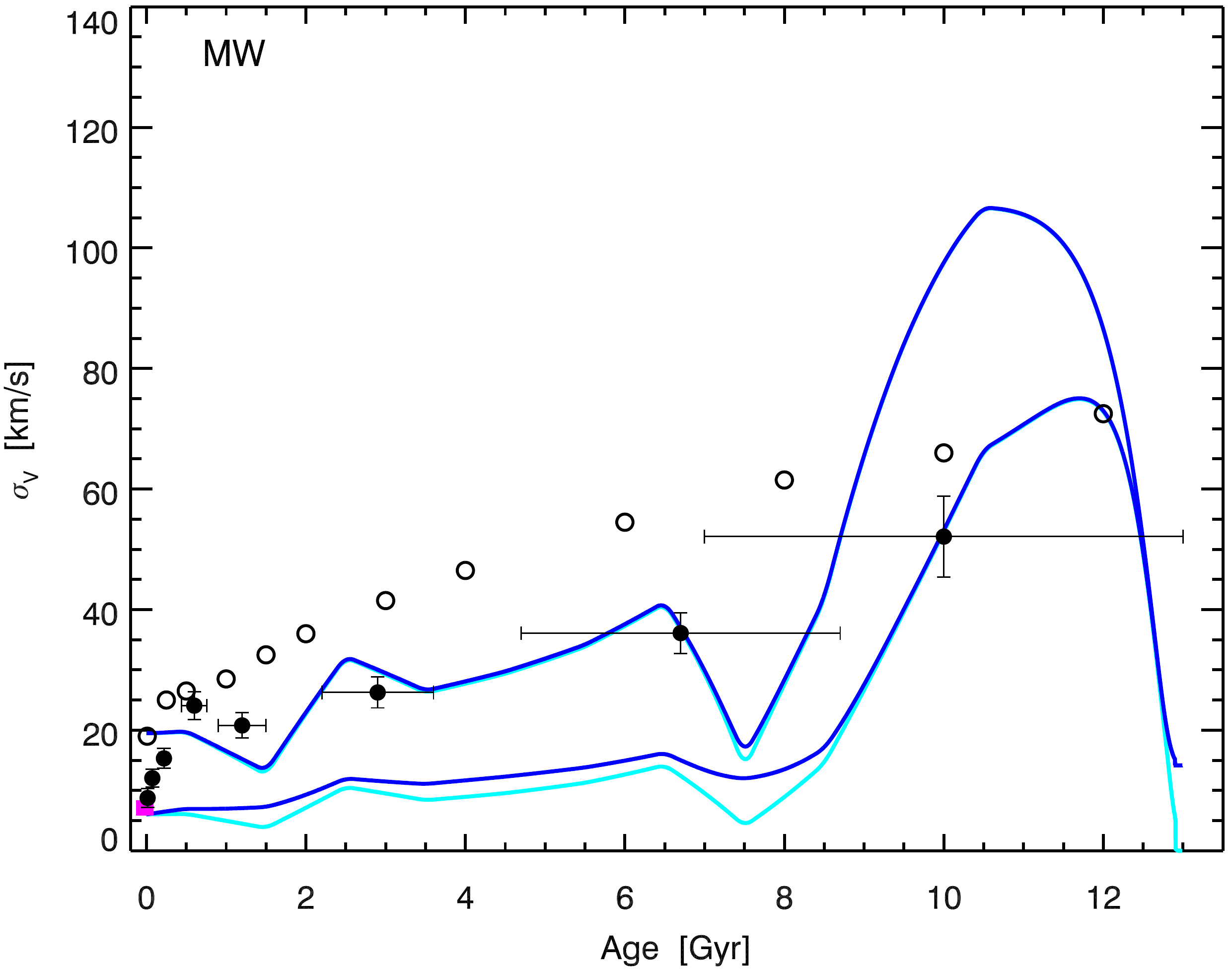}}
\subfigure{\label{fig:oneheatm31}\includegraphics[width=62mm]{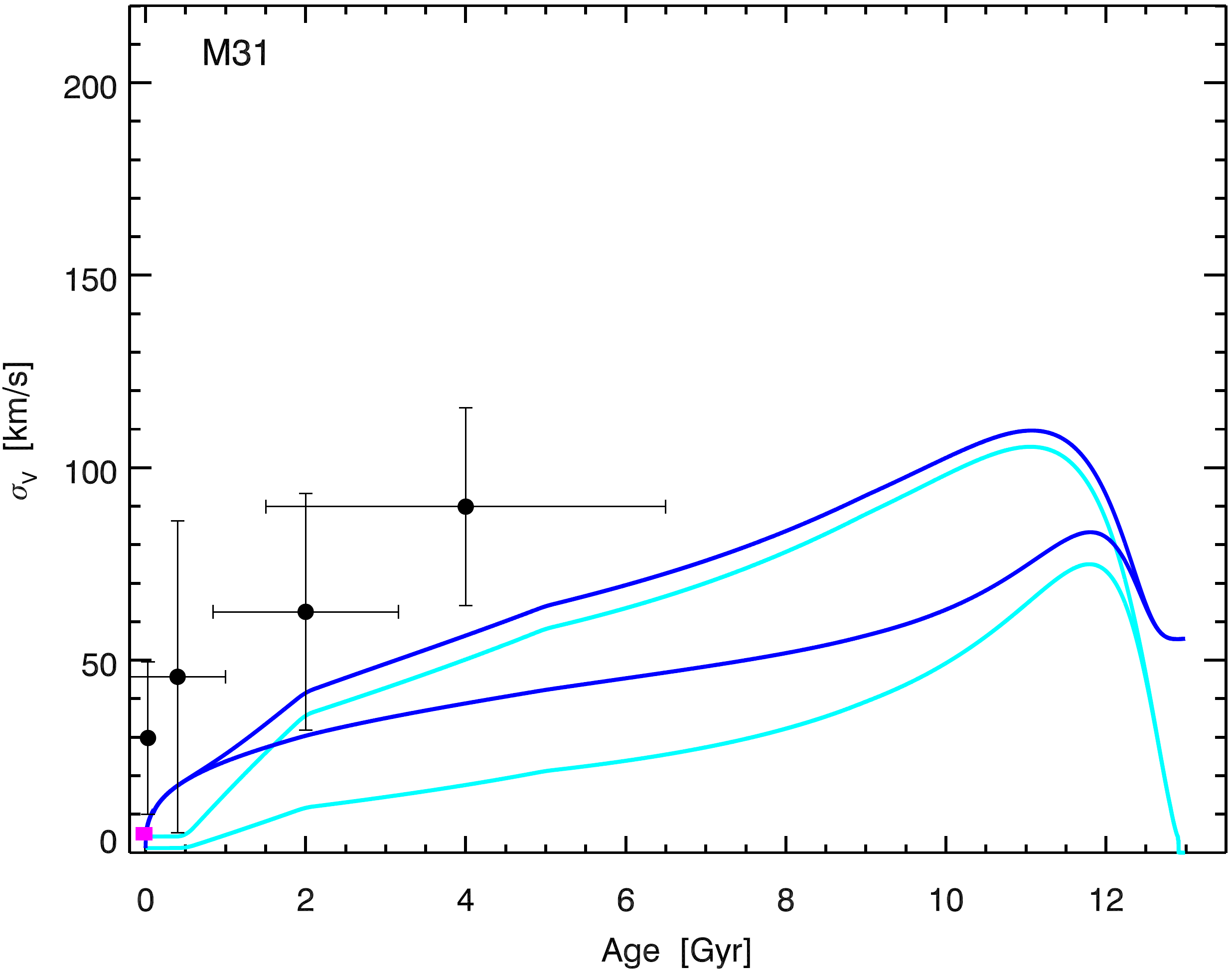}}
\caption{Stellar velocity dispersion as a function of age for the Local Group galaxies (black points).  The error bars represent the uncertainty on velocity dispersion and also account for projection and anisotropy uncertainties.  The model prediction for the $\sigma_{ISM}$ evolution is shown as the cyan lines using the observed SFH and two different gas depletion times.  The final increase in velocity dispersion of the stellar population born at $\sigma_{ISM}$ due to velocity dependent scattering off of mid-plane disk overdensities (e.g., GMCs), is shown as the corresponding dark blue lines according to Equation 6.  The magenta line shows the observed velocity dispersion of the molecular gas at present day.  For the MW both the $\sigma$ of open clusters in the disk (filled circles) and individual stars in the solar neighbourhood GCS survey are shown (open circles).  For M33 the RGB sample (open circles) of Beasley et al. (in prep) is shown relative to the star clusters (filled circles).}
\label{fig:oneheatall}
\end{figure*}

\section{Comparison of Data and Model Predictions}
The model predictions for the ISM birth velocity dispersions and the velocity dependent dynamical scattering of the stars by GMCs are shown for all the galaxies in our sample in Figure \ref{fig:oneheatall}.  The width of the $\sigma_{ISM}$ tracks represent the predictions for a molecular gas depletion time 0.3 or 1 Gyr, and are compared to observations of the present day molecular gas velocity dispersion for the SF galaxies in the sample (WLM; \citealt{Elmegreen13}, LMC; \citealt{deboer98}, M33; \citealt{Druard14}, MW; \citealt{Stark89}, M31; \citealt{CalduPrimo16}).  The calibrated diffusion coefficients are listed for each galaxy as described in the previous section.  The velocity dependence of this scattering is evident, as the process is only effective for initially cold populations ($\leq 10$km s$^{-1}$), which are susceptible to even minor perturbations, and for longer periods of time than more dynamically hot and vertically extended populations.

Importantly, we stress that the \textit{the models use only the observed SFHs as an input}, and provide a data driven insight into how the gas velocity dispersion (and therefore birth velocity dispersion of the stars) behaved as a function of time.  In most cases the final model bands can encompass the data, indicating that a time dependent $\sigma_{ISM}$ as observed at high-redshift plus internal scattering may explain a large fraction of the observed $\sigma(t)$ evolution seen in stars of different ages at present day.  The flat AVRs of the dSphs and the model over-predictions for these systems will be discussed in detail in Section 7.4

For the LMC it should be kept in mind that torques to the disk may introduce misalignments in the stellar velocity field, however such an external perturbation may be minimized when considering regions close to the bar, and are less important when we consider \emph{total} $\sigma$ rather than a single component, such as $\sigma_{z}$.  Similarly, long term tidal heating serves to only increase the lowest velocity dispersions of the LMC model ($\leq 15$ km s$^{-1}$) by $50$\% when we run impulsive tidal heating models for the LMC interaction with the MW.  Interactions with the SMC may also come into play (see Section 7.1), and it is possible that some fraction of the spectroscopic data for the LMC contain a population of accreted SMC stars which could complicate the model comparison. However given the large uncertainties on the data points our main conclusions are likely insensitive to this.

It is interesting that there is an almost universal rapid dynamical evolution for the coldest, youngest populations in the star forming galaxies in our sample (seen also in simulations; \citealt{Renaud13,Martig14,Aumer16}).  The supergiant stars typically have extremely low dispersions, comparable to the measured gas dispersions, and with rotation velocities closely mimicking the gas velocity field, but even within $\sim 1$ Gyr, the youngest RGB stars (and open clusters in the MW) become dynamically decoupled from the gas.  Given the unlikely probability that mergers occurred in this 1 Gyr time frame at low redshift for all galaxies in our sample, this dynamical decoupling is simply a manifestation of the velocity dependent mid-plane scattering discussed in Section 5.1.  The approximate mass \emph{independence} of this rapid dynamical decoupling timescale could suggest that the characteristic mass and density of the perturbing disk population is also growing accordingly.

\subsection{Leveraging Multiple MW Disk Tracer Populations}
An interesting and much studied case is the MW \citep{SS53,Wielen77,Bird13}, for which the smoothness of the AVR and any possible plateaus may provide clues to the origin (or not) of a discrete (or continuous in the case of mono-abundance populations; e.g., \citealt{Bovy12}) thick disk component(s).  We refer the reader to the detailed studies by \cite{Stinson13,Bird13} on the detailed shape of the AVR of the MW, however in this section, we wish to differentiate two dynamical MW tracer populations which may provide a consistency check on our ISM cooling models.

Before the comparison, we would draw special attention to unique biases that can arise in the case of the MW where there is a volume restricted sample of stars forming the AVR data.  Significant radial migration in the MW's disk, and/or stars on eccentric orbits can enter this sample of local Solar neighbourhood stars.  While we tune our model predictions to this particular radius of the MW, it describes the birth velocity dispersion and subsequent dynamical heating of stars that have been born in, and remain in, this volume.  Correlated changes in the components of the velocity dispersion, as well as a $\sim 10\%$ underestimate of the velocity dispersion may be expected based on simulations of these selection biases in the MW \citep{Aumer16} - however analysis and modeling of the total velocity dispersion in this work somewhat mitigates these more detailed issues that are introduced in comparing to volume restricted samples, as we shall see in Section 7.3.

In Figure \ref{fig:oneheatmw}, the ISM model predictions trace the data of the MW disk star clusters \citep{Hayes14}, but fall short of the $\sigma$ of individual Solar neighbourhood stars from the Geneva-Copenhagen study \citep{Casagrande11}.  For ages of $10-12$ Gyrs the model provides a closer agreement, however it is precisely these age bins where age errors (c.f., \citealt{Martig14,Aumer16}) will suppress the observed dispersions by $\sim 30-50\%$. The GCS stars are therefore assuredly above the model predictions even in these age bins as well. The relative difference in the two samples is unlikely to be due to contamination from halo or thick disk stars, as if anything the solar neighbourhood GCS sample of individual stars, spans a smaller survey volume than the open cluster sample, and halo contamination is estimated at $\leq 0.5$\% \citep{Cubarsi10}.  We suggest that this difference in velocity dispersion is perhaps expected and provides some reassurance that our model accurately describes the galaxies' dynamical evolution - as the star clusters should not be as susceptible to additional latent dynamical heating from spiral arms or substructures, as their mass is comparable to that of any perturbing population.  Conversely the velocity dispersions of the individual stars should be affected by any satellite mergers or spiral overdensities, and thus appear at higher $\sigma$ values at any given age.  While some studies \citep{Ida93,Ida99,Gustafsson16} have argued for comparable scattering for stars and star clusters, this is only true for mass ratios of $\mu \lesssim 1:10$.  The proportional increase in scattering for stars over star clusters increases by a factor of 1.5 if one considers even marginally higher $2:10$ events (Equations. 7-10b; \citealt{BT87}), which for the most massive open clusters forming out of a cluster and GMC mass spectrum with negative power law index is assured.  This factor increase is approximately what is seen in comparing the field star and open cluster sample in Figure \ref{fig:oneheatmw}.

\section{Discussion}
The agreement between the data and explicit model predictions indicates that we are capturing some of the most important physical processes driving the gas velocity dispersion (and initial stellar velocity dispersion, assuming the stars are born with $\sigma_{ISM}$).  In the context of this model, it suggests that gravitational disk instabilities may drive a significant proportion of the observed velocity dispersions in high-redshift galaxies and the oldest stars in Local Group galaxies, primarily due to the higher gas fractions at early times in all galaxies.

\subsection{Dynamical Evolution Trends with Mass}
Figure \ref{fig:oneheatall} indicates that low mass galaxies ($M_{*} \lesssim 10^{10}$) have $\sigma(t)$ profiles which can be explained nearly exclusively by the combination of the ISM model and the equilibrium scattering process.  Here, the intrinsically low velocity dispersions are reproduced by gravitational instabilities driven by large gas fractions, and for the least massive dwarfs the velocity dependent mid-plane scattering produces excellent agreement with the data.  For intermediate mass galaxies with $\sigma \gtrsim 15$ km s$^{-1}$, the scattering has little effect, and the ISM dispersion predictions are already sufficient to produce the magnitude and evolution of the $\sigma(t)$ profiles when we use the explicit SFHs to drive the model. 

The MW and M31 show present day stellar velocity dispersions that are in excess of the ISM and equilibrium scattering predictions (at least for $t \leq 9$ Gyr old stars).  While there is some leeway in the model, the solution would require depletion times that may be too long ($\geq 10$ Gyrs) - and as discussed in Section 6.1, the correct prediction for the MW disk star \emph{clusters} offers some confidence that the model is producing $\sigma_{ISM}$ values of the appropriate magnitude.

The mass dependence of the dispersion budget is summarized in Figure \ref{fig:sigratf}.  The top panel quantifies how prominent the mid-plane scattering is over the ISM dispersion at birth.  The efficiency of this scattering, as discussed in Section 5.1, can be understood by the velocity and mass dependence of the GMC mass and number density in Equation 6.  To illustrate this in Figure \ref{fig:sigratf} we over plot an analytic model where the molecular cloud mass and density in that equation are computed from the maximum unstable Jeans mass, $M_{J}$.  The cloud density is computed assuming that the total mass of gas is in clouds of $M_{J}$, and distributed throughout the disk size appropriate to a galaxy of that mass (see Appendix B).  This model of the mass dependent mid-plane scattering process, $\sigma_{cloud}$, in conjunction with Equation 3 (where $f_{gas}$ is taken as a function of mass following the empirical relations summarized in W15) provides a descriptive framework for the dispersion evolution in stellar systems in absence of external processes. Several simulated galaxies \citep{House11,Martig14,Christensen16}, where the birth and final stellar velocity dispersions are known are also shown and appear to exhibit a similar mass dependent behaviour in this parameter space.

The relative amount of such excess (above and beyond the $\sigma_{equil.} = \sigma_{ISM} + \sigma_{cloud}$ model) dynamical heating, in order to match the observed data, is shown in the top panel of Figure \ref{fig:sigratf}.  This Figure illustrates that at progressively higher masses the impact of additional contributions becomes necessary in order to explain the observed present day stellar AVR curves.  We discuss potential candidates for this process in the next section.


\subsection{Additional Latent Dynamical Heating Processes}
For galaxies in our sample like the MW, and M31, the prediction of the ISM model falls short of the present day stellar $\sigma(t)$ curves.  While there could be changes to the model input (larger, but unphysical gas fractions, extreme values of $Q$) that could bring these in closer agreement, it is also possible that some additional dynamical process could be responsible for the rest of the observed velocity dispersion.

\subsubsection{Heating from merging substructures}
One additional contribution to the stellar velocity dispersion is interactions between the stars and merging subhalos (not necessarily luminous; e.g., \citealt{Starkenburg15}).  There are several models and simulations available in the literature which provide analytic or numerical explorations of the dynamical heating from subhalos \citep{Benson04,Hopkins08,Helmi12,Starkenburg15}.  We defer discussion of the detailed mass spectrum and merger history of subhalos that would be required to not \emph{over predict} the $\sigma(t)$ curves to a future paper.  

While the subhalo mass function is thought to be scale free in $\Lambda$CDM, the effect of different gas fractions and initial disk structures can result in drastically different energy injections and disk responses for comparable mass ratio mergers.

\begin{figure}
\begin{center}
\includegraphics[width=0.47\textwidth]{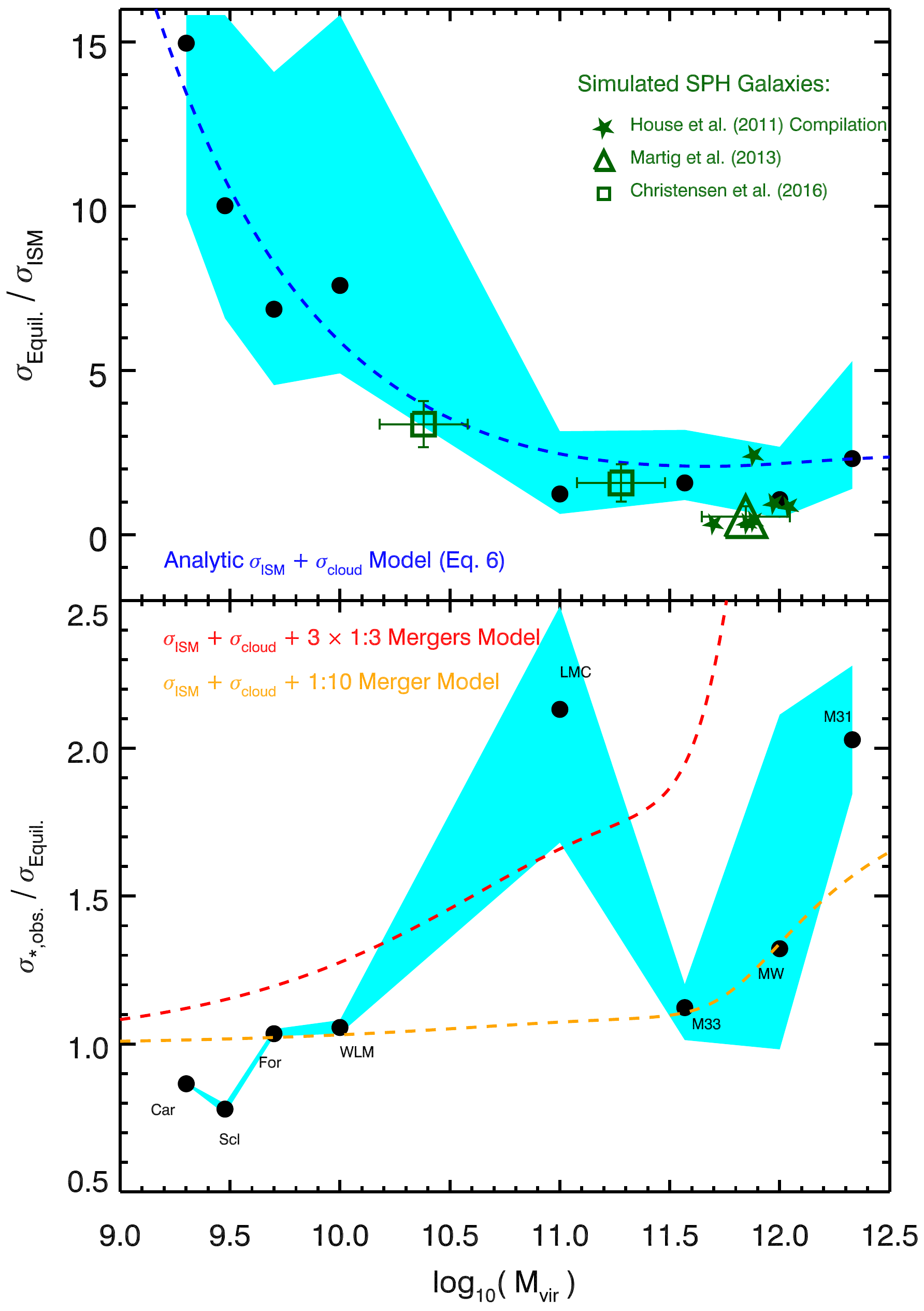}
\caption{\emph{Top:} Black points and cyan band show the ratio of the equilibrium stellar velocity dispersion ($\sigma_{equil.} = (\sigma_{ISM}^3 + \sigma_{cloud}^3)^{1/3}$; e.g., after the mid-plane scattering of stars born with the ISM gas dispersion is accounted for in the empirical SFH and scattering model).  Larger ratios for the lowest mass galaxies indicate their greater sensitivity to stars scattering off GMCs, which can also be well reproduced with an analytic scattering model (dashed blue line; Eq. 6).  Simulation data show the ratio of birth dispersion to final dispersion for various galaxies.\emph{Bottom:} Mean ratio (over all times) of the observed stellar velocity dispersion to the dispersion predicted from the ISM cooling plus mid plane equilibrium scattering model.  The higher mass galaxies show an excess velocity dispersion, which can be fit well with a model where a 1:10 merger, or repeated LMC-SMC interactions in the case of the LMC, produces additional latent dynamical heating.}
\label{fig:sigratf}
\end{center}
\end{figure}

We show one simple model, following the energy injection arguments of \cite{Hopkins08} (their Equation 8) as the dashed lines in the bottom panel of Figure \ref{fig:sigratf}.  A model that assumes that each galaxy has had a 1:10 merger, nicely reproduces the upturn in the $\sigma_{*,obs}/\sigma_{equil.}$ values, averaged over all times, at high mass.  Interestingly the LMC sits anomalously high in this parameters space - however a comfortable fit is found assuming that the increased dispersion is the result of three pericentre passages with a LMC:SMC mass-ratio companion, consistent with recently estimated orbital histories for the pair \citep{Besla15}.  While these models don't take into account the numerous complexities of how disk response evolves with scale height of the primary, they indicate that the mass dependence of the dispersion deficit may plausibly be reproduced through environmental encounters such as galaxy mergers.

\subsubsection{SN driven ISM Turbulence}
A second process which may work on top of the mid-plane scattering and gravitational instabilities discussed in this paper, is turbulence in the ISM driven by SNe feedback.  Specifically type II SNe, as SNe Ia would not be expected to be co-located in the dense molecular gas layer to the same extent as massive stars.  Recently, \cite{Lehnert14} proposed that the large velocity dispersions observed in high redshift galaxies, and in the oldest disk populations of the MW could be produced by energetic SNe feedback directly imparting turbulent motions into the gas.  They used an inferred SFH from \cite{Snaith14} to describe the SFR of the MW (and therefore energy from SNe) as a function of time, provided some coupling parameter which describes how efficient the energy injection of the SNe is.  Specifically they adopted $\sigma_{ISM} = \epsilon\sqrt{\Sigma_{SFR}}$.  

By assuming that energy from SNe was the sole source of ISM velocity dispersion, and assuming like previous studies \citep{Wielen77} that the ISM floor is always $\sigma \simeq 5-10$ km s$^{-1}$, they required a value of the coupling constant of $\epsilon = 110$ km s$^{-1}$ ($M_{\odot}$ yr$^{-1}$ kpc$^{-2}$)$^{-1/2}$.  As in \cite{SS53} and \cite{Wielen77} this methodology calibrates the efficiency of energy injection (or efficiency of scattering) so that the data are reproduced from that \emph{singular} process, with the additionally canonical assumption that the initial velocity dispersion of the gas (or stars at birth) is cold and constant at all epochs.

While these studies serve as a proof of concept for a particular process, they are limited in describing how likely it is that a mechanism is able to work in concert with other processes.  An alternative method is to test whether the energy provided by SNe over the lifetime of the galaxy is alone sufficient to power the observed turbulent energy in the gas at any given time.  Such an analysis was outlined in \cite{Stilp13}.

For a given SFR, the energy produced by the SNe is given in \cite{Stilp13} as:
\begin{equation}
E_{SF} = \eta_{SN}(SFR\times t_{diss})10^{51} {\rm ergs},
\end{equation}
where $\eta = 0.0013$ is the number of core collapse SNe per unit stellar mass for a Kroupa IMF up to 120 $M_{\odot}$.  The energy in a turbulent ISM will be dissipated over a timescale:
\begin{equation}
t_{diss} = 9.8\times10^{6}\left(\frac{\lambda}{100 {\rm pc}}\right)\left(\frac{\sigma_{ISM}}{10 {\rm km/s}}\right)^{-1} {\rm yr}
\end{equation}
The turbulent driving scale is shown to be $\lambda = 100 \pm 30$ pc from simulations of SNe driven turbulence by \cite{Joung06}.

\cite{Stilp13} define the turbulent energy budget of the ISM as $E_{turb,gas} = 3/2 M_{gas}\sigma_{ISM}^{2}\times 1.989\times 10^{43}$ ergs, where $M_{gas}$ is in solar masses.  Using these equations we can test whether the observed SFRs in our galaxies are sufficient to drive the turbulent energy of the ISM via SNe feedback.  Specifically, the limiting case of the SNe providing \textit{all} the energy to drive the observed present day dispersions (e.g., the proposal from \citealt{Lehnert14}) can be  computed by setting $\sigma_{ISM}$ in the above equations to the observed $\sigma$ of the stellar component measured at present day in our galaxies, and using the $M_{gas}(t)$ from our $t_{dep} = 1$ Gyr explicit model.

\begin{figure}
\begin{center}
\includegraphics[width=0.47\textwidth]{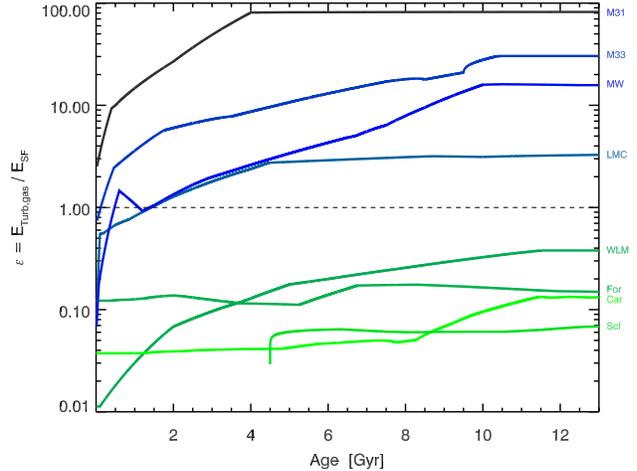}
\caption{SF Feedback efficiency parameter $\epsilon = E_{turb,gas} / E_{SF}$ as calculated from the observed SFHs and derived gas masses of the Local Group galaxies.  Tracks are coloured by stellar mass as in Figure 1.  For the galaxies above the mass of WLM, the observed $\sigma(t)$ data cannot be singularly explained with SNe driven turbulence.}
\label{fig:heatenerg}
\end{center}
\end{figure}

Figure \ref{fig:heatenerg} shows the ratio of required energy to drive the gas turbulence to energy available from SNe ($\epsilon = E_{turb,gas} / E_{SF}$).  Values of $\epsilon \geq 1$ for the high mass galaxies (LMC, M33, MW, M31) in our sample show that SNe driven turbulence is not a plausible source of energy to explain the ISM turbulence alone (for example, at least for $t \geq 1.5$ Gyrs for the MW, corresponding to $\sigma \sim 25$ km s$^{-1}$).  For WLM and the dSphs, it appears that energy from SNe could be a driver of the turbulence, particularly at late times - which may help explain the flat AVR curves of the dSphs.  We note that this exercise is using the molecular gas mass and dispersion, as opposed to the HI phase which was considered in \cite{Stilp13}, with those authours noting that there is likely a stronger coupling of the SNe energy to the denser gas phases.

This is consistent with analytic limits showing that SNe energy can not completely supply the turbulent energy budget when the gas dispersion is larger than $\sim 12-25$ km s$^{-1}$ (see Appendix A).  The mass dependent impact of such stellar feedback is seen in numerical studies where alterations to the dynamical, chemical and structural properties of low mass dwarf galaxies may result from SNe and winds of massive stars \citep{Brooks07,Governato10,Zolotov12,Teyssier13,Shen14}.  However exact ratio of energy from ionizing radiation, magnetic fields, and SNe feedback may be currently difficult to disentangle in simulations due to resolution effects.  For example if after SNe explosions the ISM cooling is turned off for a significantly longer time than the typical $t_{diss}$ timescale of $10 $Myr, then the energy from that SSP will be preserved artificially longer - while in reality perhaps constant stellar winds or magnetic field support is contributing a higher fraction of the turbulent energy budget.

The expected correlation between $\sigma_{SNe} \propto SFR$ in the case of SNe driven feedback, may therefore be strongest below $\sigma \sim 25 $km s$^{-1}$.  However more problematic in separating the contribution from SNe and gravitational sources is that, typical SF laws invoking only gravitational collapse with no feedback \emph{also} predict a correlation between the velocity dispersion and SFR.  For example \cite{KrumBurk16} have $\sigma \propto \frac{SFR}{V_{c}^{2}f_{gas}^{2}}$, which for a simple Toomre disk stability (e.g., Equation 3) leads to $\sigma \propto SFR^{1/3}$.  How best then to disentangle which contribution is dominant in our sample?

\begin{figure}
\begin{center}
\includegraphics[width=0.47\textwidth]{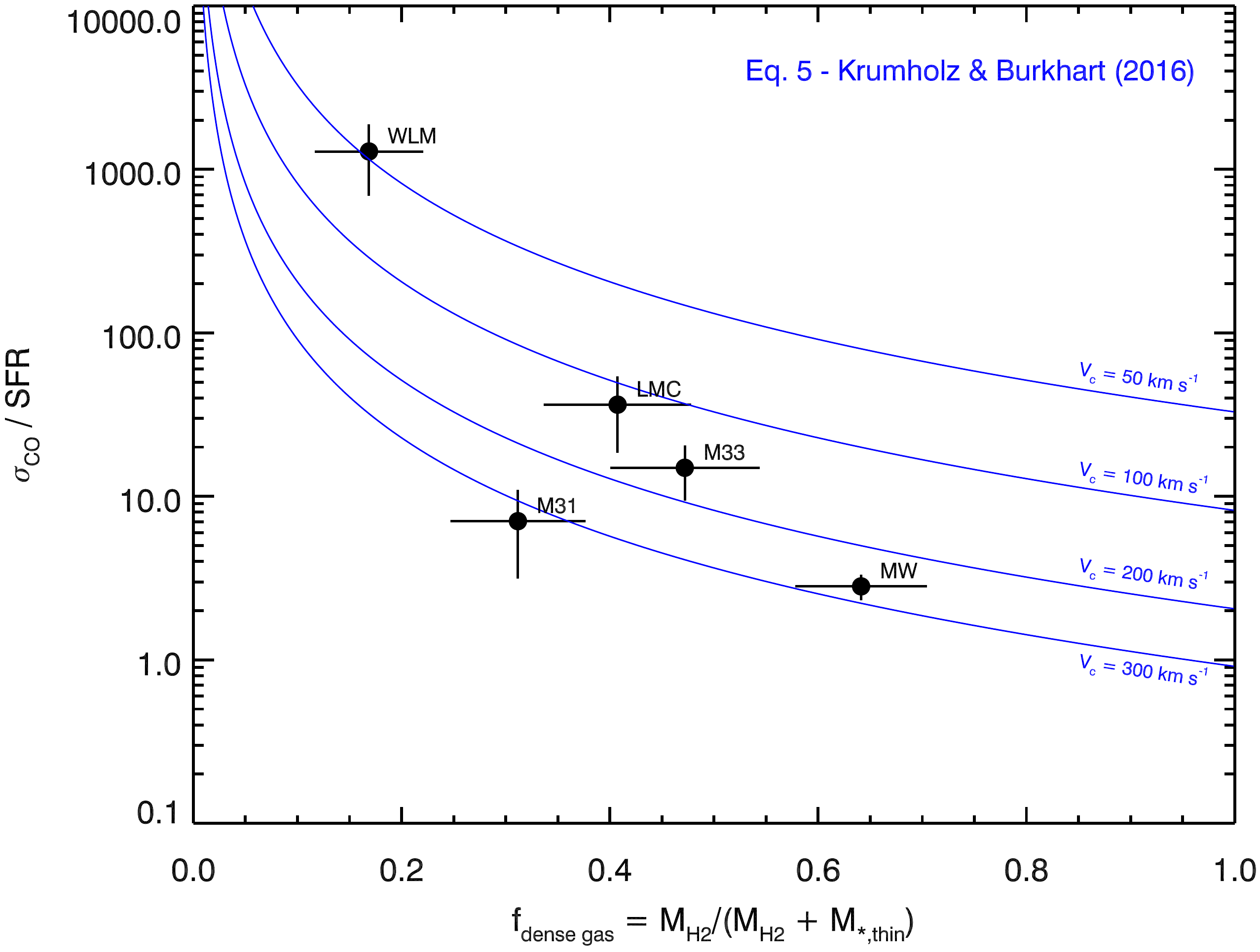}
\caption{Adapted from \protect\cite{KrumBurk16}. Ratio of \emph{observed} molecular gas velocity dispersion to SFR, as a function of present day molecular gas fraction in the five star forming galaxies in the sample.  Model lines (\emph{blue}) show the expected ratio between velocity dispersion and SFR if gravitational instabilities are the primary driver - which results in a quadratic dependence on the gas fraction.}
\label{fig:sigsfrrat}
\end{center}
\end{figure}

Recently \cite{KrumBurk16} derived expectations for the dependencies between $f_{gas}$, SFR and $\sigma_{ISM}$ where either: 1) gravitational instabilities (e.g., a Toomre criteria argument as in W15 and described in Section 5), or 2) SNe feedback were responsible for driving the turbulence and regulating SFR in the ISM.  They showed that theoretical predictions for the ratio of $\sigma_{ISM}/SFR$ diverged in the two cases - with the gravitational instability mode predicting a $\propto f_{gas}^{  -2}$ dependence, while the $\sigma_{ISM}/SFR$ was \emph{independent} of gas fraction in the SNe feedback scenario.  They compared these dependencies with observed properties in more distant galaxies, confirming the $\propto f_{gas}^{  -2}$ dependence.  In Figure \ref{fig:sigsfrrat} we reproduce this using only the \emph{observed values} of $\sigma_{ISM}$ (the present day observations of CO dispersion), SFR from the resolved SFHs and $f_{gas}$ (from the present day observed $M_{H2}$, $M_{*}$), and recover a similar $f_{gas}^{  -2}$ dependence.  While more observations of Local galaxies with comparable $V_{c}$ but differing $f_{gas}$ would be helpful, there are indications of a quadratic dependence in the data.  As with the sample of galaxies \cite{KrumBurk16} studied, these trends may point to SNe driven turbulence playing a sub-dominant role in mediating the SFR and ISM dispersion in galaxies - however here we have the added possibility of being able to test this directly on the cold-phase gas fraction and dispersion, instead of from ionized gas.

While these limiting cases exclude SNe as a singular driver of the observed stellar velocity dispersion evolution in the Local Group galaxies, it is likely present as an additional source of energy injection - albeit one with more free parameters to constrain.  On top of this the link between the $\sigma$ evolution and SFR may not be instantaneous if most energy is released into the ISM by low mass stellar SNe or ionizing radiation \citep{Peters12}.  Conversely the gravitational instabilities are intrinsic to any galactic system, and their impact on $\sigma_{ISM}$ must be present.  This process then defines a baseline level of ISM turbulence from which secondary processes (mergers, SNe energy, magnetic fields) can be calibrated against and better understood.

A discussion of the relative contribution of these additional external (mergers) and internal (stellar feedback) processes is outside the scope of this paper.  We do note however that as the disk response to mergers is approximately inversely proportional to the disk density and current velocity dispersion, a wide range of additional $\sigma_{ISM}$ contributions from SNe feedback could all be consistent with (minor) mergers still occurring in the MW disk and not overheating it. 

\subsection{Comparison to simulations}
\begin{figure}
\begin{center}
\includegraphics[width=0.47\textwidth]{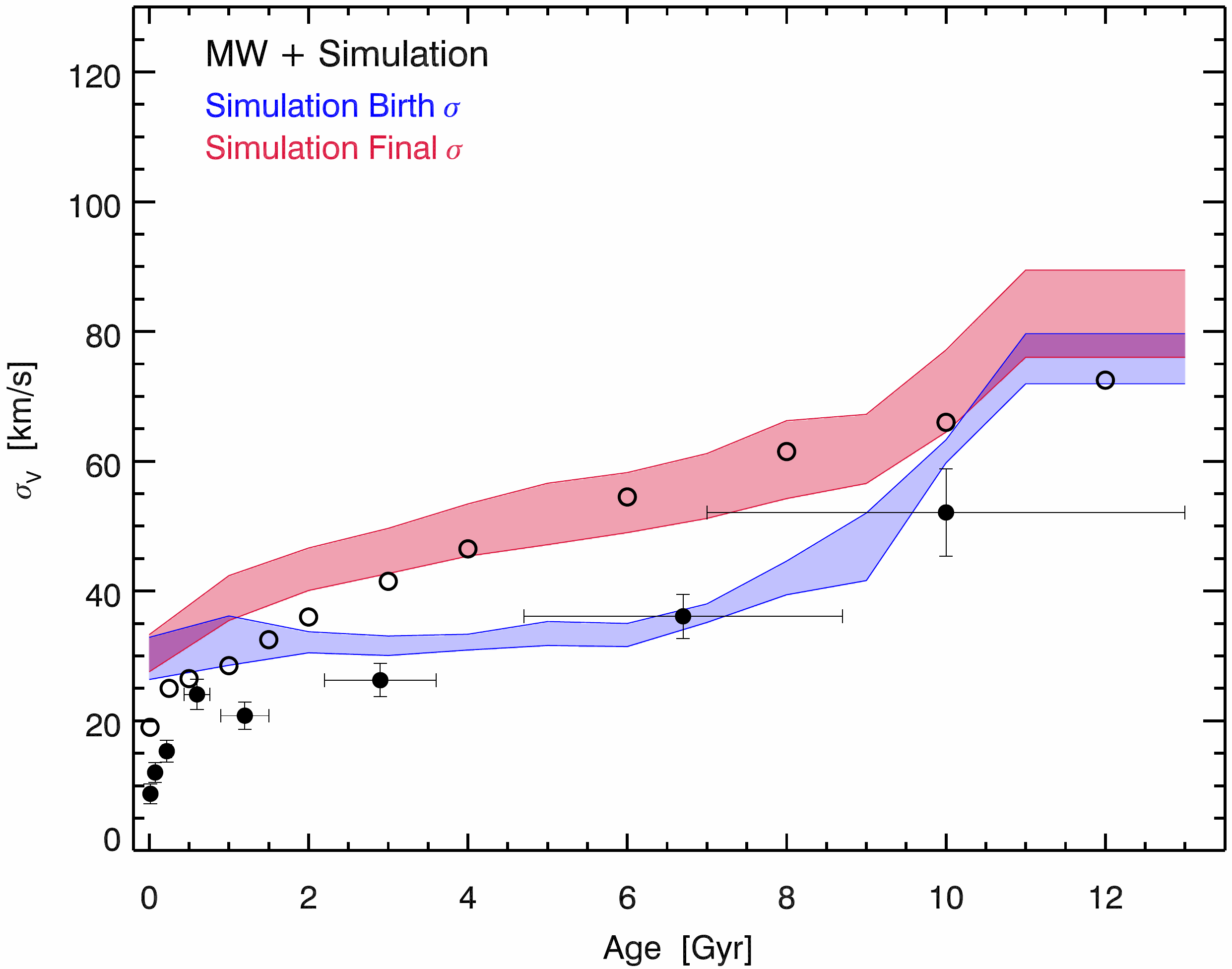}
\caption{Observed MW AVR compared with a simulated MW mass galaxy from Martig et al. (2014).  The blue and red bands are the stellar velocity dispersions at birth and at the end of the simulation.  The birth dispersions are consistent with the ISM cooling model presented in this work, and the expectations that low mass stars are dynamically heated by disk overdensities (spiral arms, GMCs) more efficiently than star clusters.  This simulation includes mergers and disk passages in addition to the mid-plane scattering effects we model, confirming also for the high mass MW that these may work in conjunction with internal effects.}
\label{fig:simheatmw}
\end{center}
\end{figure}

Another test of our ISM model prediction is to look at high resolution simulations of galaxies, where $\sigma_{ISM}, \sigma_{*,birth},$ and $\sigma_{*,final}$ can all be examined directly.  In Figure \ref{fig:simheatmw} we show the AVR curves for a simulation of a MW mass galaxy from \cite{Martig14}.  The velocity dispersion of the simulated stars at their time of birth, provide close agreement with our empirical ISM prediction, as well as the present day positions of the open cluster disk sample.  The final simulated velocity dispersions nicely track the observed $\sigma(t)$ curves of the stars, with the width of the simulation curves representing a range of radial selections (from 2 to 3 disk scale lengths) used in computing the dispersions. 

As this simulation is not an isolated galaxy, it has experienced minor mergers/disk passages, so the agreement between the redshift zero dispersion and the individual MW stars agrees with our suggestion in $\S 6.1$ that scattering effects will produce a larger effect on the individual stars, whereas the more massive open clusters should retain a dispersion closer to what they are born with.

Interestingly, similar to Figure \ref{fig:oneheatmw}, the \emph{birth} velocity dispersions are comparable to the MW disk cluster AVRs.  This gives some reassurance that our ISM model is physically motivated, as it also predicts a birth value consistent with these tracers - which should not be subject to dynamical heating to the level that individual stars are.

For the low mass galaxy regime, we compare the stellar velocity dispersion evolution of WLM to cosmological zoom-in simulations.  These simulations were computed using the smoothed particle hydrodynamic code {\sc gasoline} \citep{Wadsley04}.  This object was selected from a 25$^3$ Mpc$^3$ zoom-in cosmological simulation, and has force resolution of 87 parsecs for the $M_{vir} = 2.4 \times 10^{10} M_{\odot}$ galaxy.  Further aspects of the simulations are described in greater detail in \citep{Christensen16}.

\begin{figure}
\begin{center}
\includegraphics[width=0.47\textwidth]{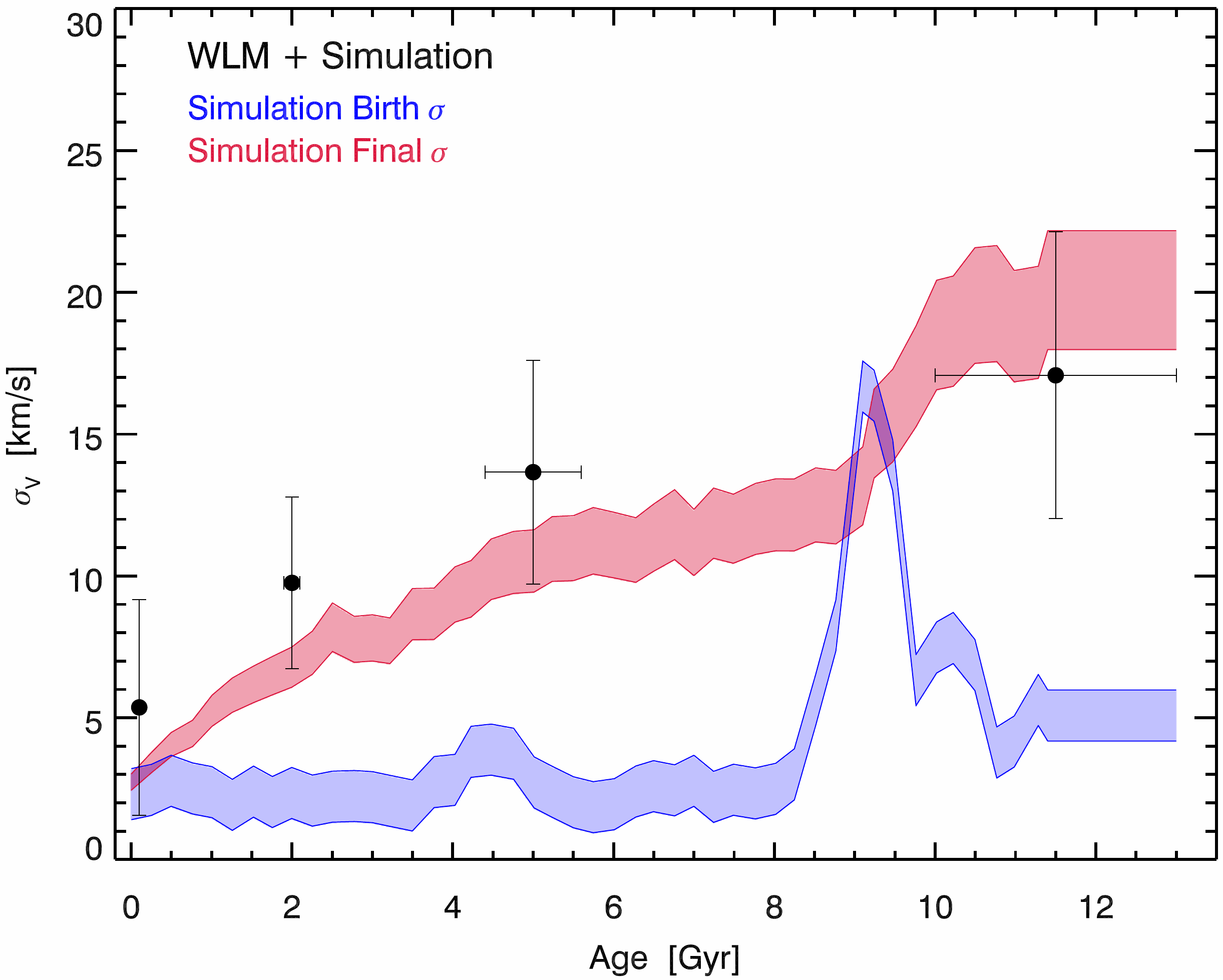}
\caption{Same as Figure \protect\ref{fig:simheatmw}, but for WLM.  Simulations overlaid are from \protect\cite{Christensen12}, and the width of the simulated dispersion tracks represents the uncertainty in converting the simulated z component of the dispersion to a total dispersion following Equation 1 and the parameters for WLM in Table 4.}
\label{fig:simheatwlm}
\end{center}
\end{figure}

Figure \ref{fig:simheatwlm} shows that the dispersion of the present day stars in the WLM-mass simulated galaxy also agrees closely with the observed AVR (see also \citealt{Teyssier13, Shen14,Brooks14}).  In addition our models' predicted ISM dispersion (see Figure \ref{fig:oneheatwlm}) is in very good agreement with the velocity dispersion of the stars at their time of birth in the simulations.

It should be kept in mind that these two simulations have different prescriptions and efficiencies for the dominant ISM feedback included in their models (SNe and stellar winds).  
For example in \cite{Christensen16} supernovae feedback is the dominant regulator of star formation in these simulations and provides one of the main sources of ISM turbulence.
These simulations use a sub-grid ``blastwave” method for including supernovae feedback \citep{Stinson06}, in order to prevent the fast dissipation of thermal energy from unresolved supernovae.  In this method the supernovae feedback is modeled by distributing energy to those gas particles within the maximum theoretical radius of the blastwave \citep{Chevalier74}, and disabling cooling in them for a time equal to the momentum-conserving phase of the supernova remnant \citep{McKee77}, typically $\sim 10^{7}$ years. Despite the temporary disabling of cooling, the particles remain hydrodynamically coupled to the rest of the simulation and rise naturally from the disk.  It is likely not trivial how the kinetic energy from the heated particles couples to, other ISM particles and if this energy is dissipated on timescales that follow $t_{diss} \propto \sigma^{-1}$.

Gravitationally-driven turbulence provides the other potential source of ISM velocity dispersion in the simulation, and it is generated spontaneously from the local gravitational and thermodynamics of the gas particles.  In order to model the full thermal range of the gas, these simulations include modeling of the non-equilibrium molecular hydrogen abundances \citep{Christensen12}, gas and dust shielding, and metal-line cooling. As such, the gas is able to reach temperatures as low as 10 K and densities greater than 100 amu/cc which should help capture many of the expected gravitational overdensities in real galactic gaseous disks.
\footnote{It should also be noted the these simulations do not include artificial pressure support against Jeans collapse \citep[][e.g.]{Robertson08}, as the high resolution of the simulations and the relatively high star formation efficiencies (10\%) mean that any gas particle that does become Jeans unstable quickly forms stars.}

In fact, one of the primary effects of including molecular hydrogen is that the ISM becomes significantly clumpier \citep{Christensen12}.  In addition to driving turbulence in the ISM through gravitational collapse, these high-density peaks may scatter stars, increasing their velocity dispersion.  While the full mass spectrum of mid-plane scattering may not be captured, the quantitative agreement between the simulations and ISM and scattering models suggest much of the large scale effects are taken into account.  In addition the recent suite of MW simulations from \cite{Grand16}, show qualitatively similar explanations for the the simulated AVR evolution.

Suffice to say that given the complexity of treating SNe feedback and its mass dependence in modern simulations, the comparison with the observational data and the analytic trends in Figure \ref{fig:sigratf} may offer a unique avenue to calibrate these prescriptions in the future.



\subsection{The flat AVR curves of the dSphs}
One interesting feature in Figure \ref{fig:normdisp} concerns the dSphs in the sample (Fornax, Sculptor, and Carina), which all show flatter $\sigma(t)$ profiles compared to the gas rich galaxies in the sample.  While metallicity subpopulations exhibit different spatial profiles and different projected radial $\sigma_{LOS}$ profiles in some dSphs, the radial averaging of stars in a given age bin necessary to achieve sufficient numbers of stars, recovers a similar average dispersion for metal rich and metal poor stars (especially given the scatter in metallicity at fixed age, which leads to more similar radial age profiles; \citealt{Battaglia06}). In addition the velocity anisotropy is thought to be a strong driver of the declining $\sigma$ as a function of radius - so when the dispersions are converted to $\sigma_{total}$, it is perhaps not surprising that the intrinsic differences in the \emph{total} dispersion between different age bins is largely constant within the significant statistical uncertainties.  So while the complex chemo-dynamics of the dSphs is deservedly the study of more detailed dynamical modeling attempts (e.g., \citealt{Amorisco12,Zhu16}) and future studies with larger sample size and precision will benefit from considering radial AVR trends, we can still ask in the context of our homogenized $\sigma_{total} (t)$ measurements, what implications the suggested flat AVR of the dSphs physically implies for their evolution.

Given the close proximity of the dSphs to the MW, it is worth considering whether this difference results from an environmental effect.  This is particularly worth exploring, given that a constant $\sigma_{ISM}$ evolution in these galaxies would be in disagreement with the observed SFHs and evolving gas fractions seen in simulations and high redshift galaxies - though this may speak to an environmentally modulated early star formation efficiency/history such as proposed by \cite{Gallart15}.

While some tidal stripping of the DM halos of dwarf galaxies in simulations is thought to be necessary to reconcile the central densities and velocities with observations (especially in conjunction with baryonic feedback in the most luminous dwarfs; \citealt{Zolotov12,Brooks14}), the minimal scatter in the dwarf galaxy stellar mass-metallicity relation \citep{Kirby13}, and orbital simulations \citep{Battaglia15} indicates that there has likely not been significant \textit{stripping} of the stars in these objects.  However the MW may still provide a source of tidal heating to the stars bound within the dwarf spheroidals (dSphs), potentially increasing the velocity dispersion (albeit temporarily as the dispersion would decrease after the system expands and re-virializes due to the negative heat capacity of galaxies; \citealt{Aguilar85,Aguilar86}).  

We tested impulsive tidal heating models and found that while the enhancement in dispersion due to repeated pericentre passages wouldn't flatten the $\sigma(t)$ profiles, the subsequent re-virialization of the system should result in the existing stars having a larger radius and lower velocity dispersion prior to the galaxy's pericentre encounter (provided the crossing time is less than the orbital time around the MW) \citep{Penarrubia09,Smith13,Barber15}.  If such pericentre heating-relaxation cycles happened one or more times prior to the youngest generation of stars having formed in the dSphs, it could effectively lower the relative dispersion of the older stars and produce a flatter dispersion profile.  This could naturally explain the more concentrated profiles of the metal rich components \citep{Tolstoy04} and steeper metallicity gradients in some dSphs \citep{Battaglia11,Leaman13,Ho15, Kacharov17}.

The same decrease in velocity dispersion may be expected for relaxation after potential reconfiguration due to feedback induced dark matter``cusp-core'' transformations seen in simulations of low mass dwarf galaxies \citep{PG12,dicintio14,Read17}, or due to GMC based scouring \citep{Nipoti15}.  Indeed \cite{Read17} show not only is the core creation occurring in galaxies in the mass range of the dSphs, but that the velocity dispersion in their simulated cored galaxy may be lower by a factor of two than in the cuspy case.  Given the numerous simulation studies on feedback in dwarf galaxiess, one could speculate that susceptibility of these dSphs to stellar feedback may potentially give rise to flat AVR curves as well.


An alternative or complementary mechanism could be a scenario where the inner most gas reservoir of the dSphs is shocked during ram pressure stripping, inducing the last episode of star formation.  This would provide a higher $\sigma_{ISM}$, and therefore higher $\sigma_{*}$ for the youngest populations than predicted by the current model.  Like the above heating/relaxation scenario, it would also be consistent with steep metallicity gradients in dwarf galaxies with low ratios of $V_{rot}/\sigma_{*}$ \citep{Schroyen11,Leaman13}.  While \cite{Wheeler16} have recently found similar $V_{rot}/\sigma_{*}$ for a large sample of Local Group dIrrs and dSphs when examining RGB stars of all ages, the data here suggest that the dSphs uniquely have a low $V_{rot}/\sigma_{*}$ for their \textit{youngest} stars, relative to stars of the same age.  These signatures, and the associated radial metallicity profiles could plausibly be generated in both the tidal heating or ram pressure scenarios - though further numerical study would be helpful.



A third process to consider is that the dSphs were a product of dwarf-dwarf mergers, or pre-processing in a group-infall scenario \citep{Yozin12,Yozin15,Wetzel15}.  The flat $\sigma(t)$ profiles in this case may be a natural consequence of this mechanism, whereby repeated group interactions maintains some equilibrium velocity dispersion for the gas and stars over a significant portion of the dwarf's lifetime.  The galaxy mass dependence of such pre-processing is seen in the simulations of \cite{Wetzel15}, where a much higher percentage of dwarfs with $M_{*} \leq 10^{7}$ were found to be satellites of smaller groups, before and during infall to the MW/M31 primary halos.  If such flat $\sigma(t)$ profiles are a result of group pre-processing, this would provide a natural explanation for why it is seen most clearly in the lower mass dSphs.  Additionally, recent simulations by \cite{Benitez15} indicate that this process can also leave metallicity gradients in dSphs.


\section{Conclusions}
We present comparative stellar velocity dispersions and ages from resolved spectroscopic observations of individual stars in eight galaxies in the Local Group -  ranging in mass from the low mass dSph Carina ($M_{*} \sim 10^{6} M_{\odot}$) to M31 ($M_{*} \sim 10^{11} M_{\odot}$).  Motivated by simulations and observations of increased gas velocity dispersion in high redshift galaxies \citep{Kassin12,Kassin14,Wisnioski15}, we utilize the observed SFHs from CMD analyses to estimate the gravitational disk instabilities due to varying gas fractions, and predict the $\sigma_{ISM}$ of the galaxies as a function of time.  This analysis relaxes the common assumption in AVR studies, that stars at all ages are born from cold gas disks with constant velocity dispersion, and lets us test whether stars can be born from with the dispersion of the gas disk at any epoch in a physically motivated evolutionary picture.

Application of these models, plus empirical dynamical scattering prescriptions from mid-plane overdensities (GMCs), consistently reproduces the AVRs of the lowest mass galaxies - suggesting that stars are born from gas which becomes gradually less turbulent over time due to declining gas fractions and a smaller amount of gravitationally driven turbulence in the ISM.

The exception are the three dSphs we consider, which show shallower AVRs even over more than 8 Gyrs in evolution.  Possible scenarios (all consistent with these dSphs showing steep stellar population gradients) are that the flat AVRs may be due to: increased stellar feedback or ram-pressure induced shocks prior to the youngest episode of SF, tidal heating and subsequent expansion of the older population during the dSphs orbits, or group pre-processing/dwarf-dwarf merging occurring predominantly in the evolution of these low mass dSphs.

Above masses of $M_{*} \sim 10^{10}$ additional sources of latent dynamical heating, such as mergers of subhalos are needed to reproduce the present day $\sigma(t)$ curves in M33, and especially the MW and M31.  This is plausibly accounted for by disk passages or minor mergers of satellites, for which the merger rate at fix mass ratio should increase proportionally more for such massive disk galaxies.

While the SFH based model supports the picture where stars are born from high velocity dispersion gas at high redshift - the question of what is driving the turbulence still appears complex.  The models of gravitationally driven instabilities we consider can be supplemented by secondary energy sources such as feedback from SNe or magnetic fields.  However using the SFHs and analytic arguments the energy provided by SNe feedback appears to be self-limiting and insufficient to completely drive the ISM turbulence above $\sigma \sim 20$ km s$^{-1}$, and fails to reproduce the stellar AVR curves of the high mass galaxies in this sample.  Additionally, comparing the present day SFRs, gas fractions, and molecular gas velocity dispersions, the ratio of $\sigma/SFR$ shows a dependence on gas fraction which is uniquely expected in the case of gravitational driven turbulence - strongly hinting that the high dispersions are originating from gravitational disk instabilities.

The models we consider are free of tunable parameters aside from an assumed $t_{dep}$ - yet extreme variations in this parameter do not alter the main conclusions of our results.  Furthermore we demonstrate through comparison with simulations and two distinct dynamical tracers in the MW (open clusters and individual stars) that our semi-analytic models correctly predict the relative magnitude difference for the AVR of the MW disk open clusters, relative to individual solar neighbourhood stars.  This is an excellent consistency check given that the more massive open clusters should not show as significant latent heating from feedback or mergers, compared to the stars, due to the former's larger masses. 

Taken together this strengthens a hypothesis that, in line with high redshift observations, the increased velocity dispersion of the gas at early times is driven by the galaxies' large gas fractions which induce gravitational instabilities in the ISM.  Our analysis of Local Group galaxy SFHs and AVRs suggests that the stars were likely born with dispersions close to the gas velocity dispersion at their formation epoch, and therefore offer a useful record of the dynamical history of the galaxy gaseous and stellar disk to present day.

The scaling relations and model comparisons presented here should be helpful to further refine feedback and SF prescriptions in simulations and interpretations of high redshift integrated light studies, as well as calibrate the efficiency of secondary dynamical heating processes.

\section*{Acknowledgments}
We thank the anonymous referee for many helpful suggestions which improved this manuscript.  We would like to thank Mark Gieles, Brad Gibson, Glenn van de Ven, Jonathan Bird, Scott Tremaine, Joss Bland-Hawthorn, Sarah Sadavoy and Bruce Elmegreen for useful discussions which helped improve this manuscript.  This work was supported by Sonderforschungsbereich SFB 881 "The Milky Way System" (subproject A7 and A8) of the Deutsche Forschungsgemeinschaft (DFG), and was also made possible with funding from the Natural Sciences and Engineering Research Council of Canada PDF award.  ES gratefully acknowledges funding by the Emmy Noether program from the DFG. GB acknowledges financial support by the Spanish Ministry of Economy and Competitiveness (MINECO) under the Ramon y Cajal Programme (RYC-2012-11537). MB acknowledges support from the Programa Nacional de Astronom\'{i}a y Astrof\'{i}sica of MINECO, under grant AYA2013-48226-C3-1-P

\bibliographystyle{mn2e}
\bibliography{refdynheat}

\appendix
\section{SNe Turbulent Energy Injection Limits}
While Equations 7 and 8 of this work provide a framework to evaluate the amount of energy produced by SNe for an arbitrary SFH relative to the turbulent energy capacity of the gas - here we briefly explore a potentially more fundamental limit on increase of velocity dispersion that can be achieved by SNe - independent of the galaxy's SFR. 

The dissipation timescale in Equation 8 is expressed more fundamentally in \cite{Maclow99}, who showed that the ratio of the dissipation timescale to the free-fall time of the star forming molecular cloud is inversely proportional to the cloud's Mach number ($\mathcal{M}$) and proportional to the ratio of the turbulent driving wavelength and the Jeans length:
\begin{equation}
\frac{t_{diss}}{t_{ff}} \simeq \frac{3.9}{\mathcal{M}}\left(\frac{\lambda_{D}}{\lambda_{J}}\right).
\end{equation}
As the driving wavelength of the turbulence $\lambda_{D}$ can not be larger than the actual dimensions of the collapsing cloud, $\lambda_{D}/\lambda_{J} \leq 1$ in all cases, and sets a fundamental limit for the SNe energy budget.

The Mach number depends on the velocity dispersion and the local sound speed of cold molecular gas ($c_{s} = 0.2 {\rm km s}^{-1}$) in the cloud so that the limiting case becomes:
\begin{equation}
\frac{t_{diss}}{t_{ff}} \lesssim \frac{3.9c_{s}}{\sigma}.
\end{equation}

\cite{Maclow99} noted that turbulence will only be successful at preventing collapse if the velocity dispersion is quite low (see also \citealt{Forbes16,Kortgen16}).  In cases where the gas dispersion is intrinsically high, the dissipation timescale of the turbulent energy will be much less than the free-fall time (as typical Mach numbers are 10-100 in GMCs).

We now exploit the velocity dispersion dependence of the above ratio to incorporate this limiting behaviour into the ratio of available energy from SNe relative to the turbulent energy capacity of the gas discussed in $\S 7.2.2$.
\begin{equation}
\frac{E_{SF}}{E_{gas}} = \frac{\eta SFR ~t_{diss} ~10^{51} ~{\rm erg}}{3 M_{gas} ~\sigma^{2} ~10^{43} ~{\rm erg}}.
\end{equation}

Following a Kennicutt-Schmidt type SF law, the gas mass is often expressed in terms of the free-fall time and the efficiency per free fall time of gas to star conversion (e.g, \citealt{Krumholz10}):
\begin{equation}
SFR = \frac{\epsilon_{ff}}{t_{ff}}M_{gas}.
\end{equation}

Substituting this for $M_{gas}$ in Equation 6, yields:
\begin{equation}
\frac{E_{SF}}{E_{gas}} = \frac{\epsilon_{ff}\eta~ 10^{8}}{3\sigma^{2}}\left(\frac{t_{diss}}{t_{ff}}\right).
\end{equation}

Finally from the physical limit imposed by $\lambda_{D} \leq \lambda_{J}$, we can write the upper limit for the contribution to the turbulent energy capacity of the gas from SNe energy injection:
\begin{equation}
\frac{E_{SF}}{E_{gas}} =\left(\frac{\epsilon_{ff}\eta c_{s}}{\sigma^{3}}\right)~1.3\times10^{8}.
\end{equation} 

The SFR dependency cancels, leaving only a dependence on the gas velocity dispersion and constants of the molecular cloud properties.  Observationally, compilations such as \cite{KrumTan07} have found a remarkably universal efficiency per free fall time across many galaxies and gas density tracers so that $0.006 \leq \epsilon_{ff} \leq 0.03$ brackets the range, with a median of $\epsilon_{ff} = 0.01$.  For a Kroupa IMF $\eta_{K} = 1.3\times10^{-2}$ and for a Salpeter $\eta_{S} = 2\times10^{-3}$.  Using the canonical sound speed of $c_{s} = 0.02$ km s$^{-1}$ for molecular gas, we plot in Figure 1 the limiting dispersion which SNe energy could fully account for the turbulence in the gas, for various choices of the IMF and star formation efficiency.  Even with the largest observable values of $\epsilon_{ff}$, the energy injection is insufficient to power the dispersions seen in highly turbulent gaseous disks at high redshift or in the oldest stars in MW mass galaxies.  Analytically the range of upper limits which SNe energy can provide from Equation A6 is:
\begin{equation}
11.6 ~{\rm km~ s}^{-1} \leq \sigma_{lim} \leq 24.4 ~{\rm km~ s}^{-1}
\end{equation}

\begin{figure}
\begin{center}
\includegraphics[width=0.47\textwidth]{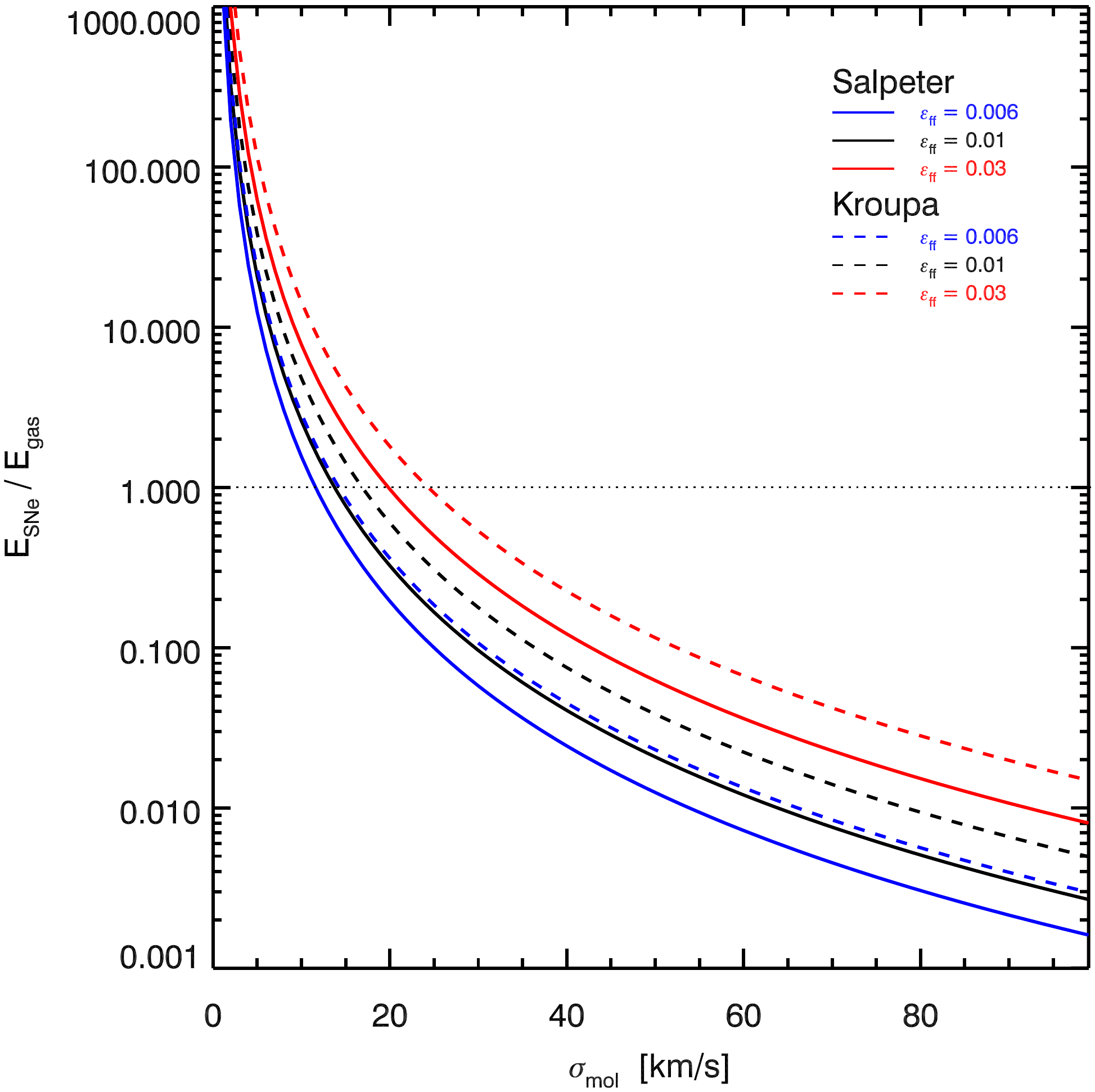}
\caption{Coloured curves show the maximum energy injected by SNe relative to the turbulent capacity of the gas, as a function of the ISM velocity dispersion.  The curves are labeled for different assumptions of the IMF and SF efficiencies.}
\label{fig:snelim}
\end{center}
\end{figure}

Figure \ref{fig:snelim} shows curves of the maximum fraction of the turbulent energy budget that can be driven by SNe for an ISM with velocity dispersion $\sigma$, assuming different choices of the IMF and star formation efficiency.  While magnetic fields, and ionizing radiation from young massive stars assuredly contribute to the gas turbulent energy budget as well, the SFR independence of this analytic limit for SNe, may naturally lead to mass dependence for the type of non-adiabatic transformations of the potential due to rapid SNe driven gas expulsion that are seen to transform cosmological DM density profiles in simulations (e.g., \citealt{PG12,dicintio14}).

\section{Analytic Mass Dependent ISM Cooling And GMC Scattering Model}
In Figure \ref{fig:sigratf} the empirical SFH based ISM cooling prediction and the empirical mid-plane scattering (e.g., the cyan band and black points) are compared to a semi-analytic galaxy mass dependent model (blue dashed line).  The ingredients for the $\sigma_{ISM,SAM}$ and $\sigma_{cloud,SAM}$ model are outlined below.

The gas velocity dispersion as a function of galaxy mass is estimated following the theoretical disk instability arguments from W15, in conjunction with the empirical trends of gas fraction as a function of galaxy stellar mass, $M_{*}$ from \cite{Whitaker14}.  Specifically at redshift zero,
\begin{eqnarray}
f_{gas} &=& (1 + (t_{dep}sSFR)^{-1.})^{-1}\\
t_{dep} &=& 1.5 ~~ {\rm [Gyr]}\\
sSFR &=& \frac{-10.73 + 1.26}{1 + {\rm exp}\left((10.49 - {\rm log_{10}}(M_{*}))/-0.25\right)}
\end{eqnarray}

The circular velocity of the halo is computed via the virial mass mapped using a stellar to halo mass relation (SHMR; here from \citealt{Leauthaud12}).  The ratio of the galaxy effective radius, and virial radius can be linked via an angular momentum conservation argument to the SHMR scaled to the same fractional radius ($\eta$) of the galaxy via derivations presented in Leaman et al. (in prep) and Diaz Garcia (in prep).  This coupled with the empirical galaxy size-mass relation (e.g., \citealt{Dutton11}) yields:
\begin{eqnarray}
\eta &=& \left(\frac{M_{vir}}{M_{*}}\right)_{\alpha R_{e}} = 0.04~{\rm exp}(0.38\alpha) - 0.04\\
R_{e} &=& 10^{0.51 + 0.2(log_{10}(M_{*}) - 10.17)}  ~~[{\rm Kpc}]\\
R_{vir} &=& \frac{\alpha R_{e}}{\eta}\\
\sigma_{ISM,SAM} &=& \frac{Q}{a}f_{gas}\left(\frac{GM_{*}}{\alpha R_{e}}\right)
\end{eqnarray}

The analytic cloud scattering model must specify the typical maximum mass ($M_{GMC,max}$) and number density ($n_{GMC}$) of the GMC population as a function of the host galaxy mass (e.g., Equation 6).  The gas mass can be trivially extracted from the empirical $f_{gas}$ relation as a function of galaxy mass (Equations B1-B3).  The maximum GMC mass will be close to the Jeans mass for the galaxy which can be expressed \citep{Genzel11} in terms of the galaxy mass and gas fraction as.
\begin{equation}
M_{GMC,max} = 3\times10^{7}\left(\frac{M_{gas}}{10^{9}}\right)\left(\frac{f_{gas}}{0.2}\right)^{2}
\end{equation}

For the limiting case where the most important perturbers have mass $M_{GMC,max}$, and the gas mass is primarily locked up in these structures, the total number of GMCs is $N_{GMC} = M_{gas}/M_{GMC,max}$.  To compute their spatial number density we assume that they are distributed within an exponential disk (with size proportional to stellar mass of the galaxy from Equation B5), and a mass dependent vertical thickness $q_{0} = h_{z}/R_{d}$ fit to the empirical relations of \cite{Kregel02}.  This yields:
\begin{eqnarray}
q_{0} &=& ({\rm exp}(0.01V_{c})+4)^{-1}\\
n_{GMC} &=& \frac{N_{GMC}}{R_{e}}\frac{{\rm exp}(-\alpha)}{2\pi\alpha q_{0}R_{e}^{2}}
\end{eqnarray}
Equation 6 then specifies that the scattering coefficient and equilibrium velocity dispersion due to scattering on top of the ISM dispersion is:
\begin{eqnarray}
\gamma_{SAM} &=& \frac{\rho_{GMC}M_{GMC,max}}{4404.83}  ~~[{\rm km^{3} ~s}^{-3} {\rm yr}^{-1}]\\
\sigma_{equil,SAM} &=& \left(\sigma_{ISM,SAM}^{3} + \frac{3}{2}\gamma_{SAM}t_{dyn,disk}\right)^{1/3}\\
\end{eqnarray}

\begin{figure}
\begin{center}
\includegraphics[width=0.47\textwidth]{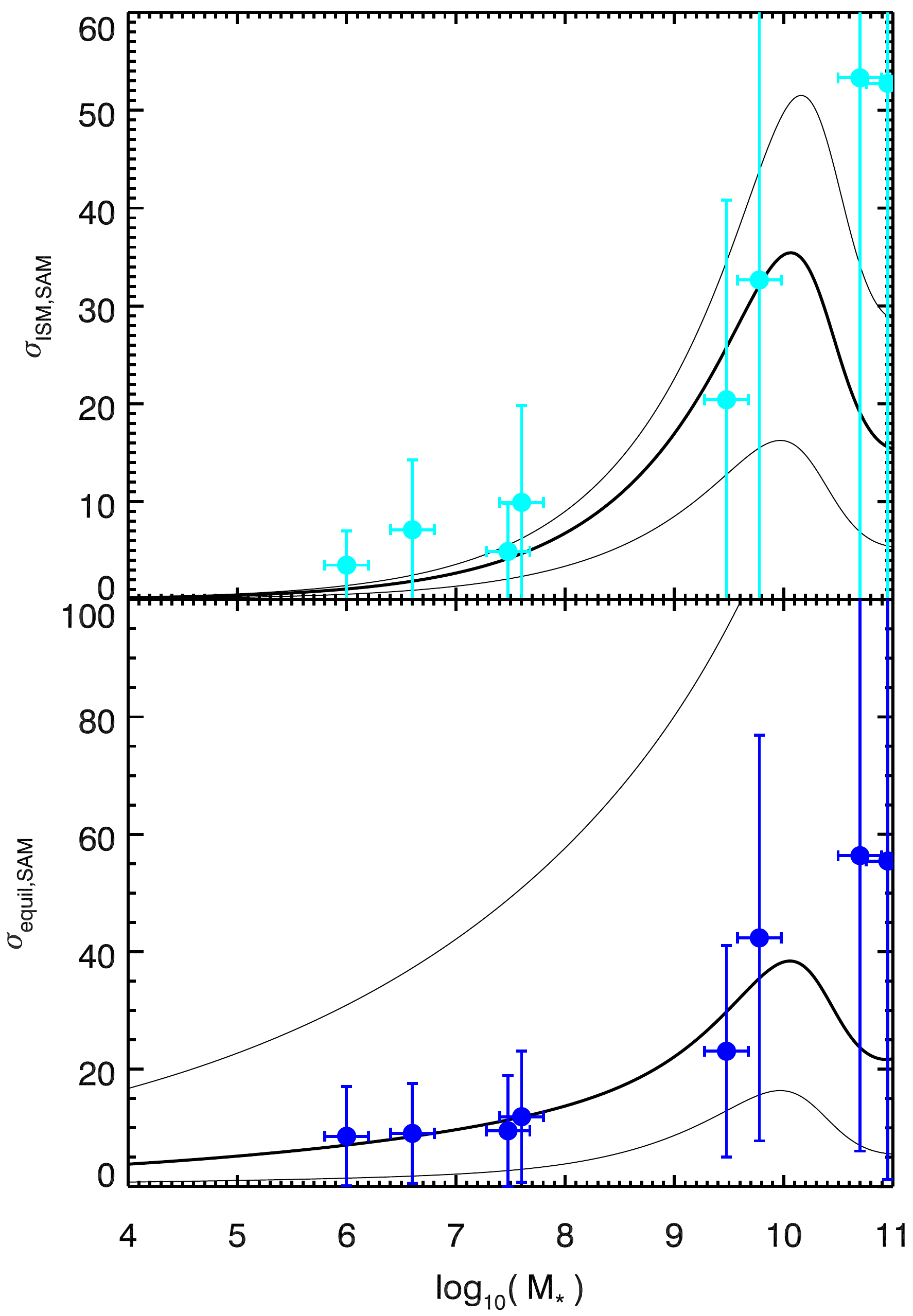}
\caption{Galaxy mass dependent semi-analytic ISM and equilibrium scattering models.  Curves show the uncertainties due to gas depletion time in Equations B1-B13, and the points and error bars show the average and \emph{range} of the velocity dispersion across all times in the \emph{empirical} ISM and mid-plane scattering models for each of the Local Group galaxies in our sample (e.g., the cyan and blue curves in Figure \ref{fig:oneheatall}).}
\label{fig:sigsammod}
\end{center}
\end{figure}

Figure \ref{fig:sigsammod} shows these analytic models for ISM dispersion, and ISM plus scattering (e.g., $\sigma_{equil,SAM}$).  The uncertainties on the models assume a variation in depletion time between $1-3$ Gyrs in Equation B2, and the 1-$\sigma$ uncertainties on the \cite{Leauthaud12} SHMR.

\bsp

\label{lastpage}

\end{document}